\documentclass[%
 reprint,
amsmath,amssymb,
aps,prd
]{revtex4-2}

\usepackage{dcolumn}
\usepackage{graphicx}
\usepackage{bm}
\usepackage{booktabs}
\usepackage{amsmath}
\usepackage{subfigure}
\usepackage[hidelinks]{hyperref}
\hypersetup{
    colorlinks=true,
    linkcolor=blue,
    filecolor=gray,      
    urlcolor=blue,
    citecolor=blue,
}

\begin{document}

\preprint{APS/123-QED}

\title{\textbf{The Gravitational Wave Forms of Galactic Compact Binaries \\
with Mass-Transfer Correction}}

\author{Zi-han Zhang$^1$}
 \email{zhangzihan19@lzu.edu.cn}

\author{Bin Liu$^1$}%
\email{liubin2020@lzu.edu.cn}

\author{Sheng-hua Yu$^{2}$}
\email{shenghuayu@bao.ac.cn}

\author{Jie Yang$^{1,3,4,5}$}
\email{Corresponding author: yangjiev@lzu.edu.cn}

\affiliation{$^1$School of Physical Science and Technology, Lanzhou University, Lanzhou 730000, China}%
\affiliation{$^2$CAS Key Laboratory of FAST, National Astronomical Observatories, Chinese Academy of Sciences, 20A Datun Road, Beijing 100101, China}
\affiliation{$^3$Institute of Theoretical Physics $\&$ Research Center of Gravitation, Lanzhou University, Lanzhou 730000, China}%
\affiliation{$^4$Key Laboratory of Quantum Theory and Applications of MoE, Lanzhou University, Lanzhou 730000, China}%
\affiliation{$^5$Lanzhou Center for Theoretical Physics $\&$ Key Laboratory of Theoretical Physics of Gansu Province, Lanzhou University, Lanzhou 730000, China}%


\date{\today}

\begin{abstract}
In this paper, we focus on the effect of mass-transfer between compact binaries like neutron-star—neutron-star (NS-NS) systems and neutron-star—white-dwarf (NS-WD) systems on gravitational waves (GWs). We adopt the mass quadrupole formula with  2.5 order Post-Newtonian (2.5 PN) approximation to calculate the GW radiation and the orbital evolution. After a reasonable discussion of astrophysical processes concerning our scenario, two kinds of mass-transfer models are applied here. One is the mass overflow of the atmosphere, where the companion star orbits into the primary's Roche limit and its atmosphere overflows into the common envelope. The other one is the tidal disruption of the core, which is viewed as incompressible fluid towards the primary star, and in the near region branches into an accretion disc (AD) and direct accretion flow. Viewing this envelope and as a background, the GW of its spin can be calculated as a rotating non-spherically symmetric star. We eventually obtained the corrected gravitational waveform (GWF) templates for different initial states in the inspiral phase.
\vspace{0.4in}
\end{abstract}

\maketitle


\section{INTRODUCTION} 

The galactic compact binaries are important sources of celestial GW detection \cite{PhysRevD.79.062002,Liu_2022}. They are within the sensitive frequency band (0.1 mHz $\thicksim$ 0.1 Hz) of detectors and they are rich in number \cite{PhysRevD.100.064060}. GW signals from these sources contain a wealth of information about their formation and evolution, mass transfer, and the equation of state \cite{PhysRevD.85.122005}. Establishing an accurate gravitational waveform (GWF) template library for close compact binary stars with Roche lobe overflow, material exchange, and other influences can not only help to analyze waveforms but also benefit data processing for numerous compact binaries. Along with that, when searching for other outer galactic objects we are interested in, noises from inner compact binaries can be reduced \cite{PhysRevD.106.102004,2006A,2007A,PhysRevD.85.127502,PhysRevD.79.044030}.

The strain and frequency of GWs vary from different compact gravitational wave sources, which commonly include BH-BH, BH-NS, NS-NS, NS-WD, and WD-WD  \cite{Wagg_2022}. The GWs of Extreme Mass-Ratio Inspirals (EMRI) and Compact Binary Coalescence (CBC) are research hotspots of this field and have been widely studied. The current terrestrial detectors like LIGO and Virgo, along with celestial detectors like LISA, TaiJi, TianQin, and DECIGO \cite{Kang_2021,Fan_2022,10.1093/ptep/ptab019}, have different sensitivity in the strain and frequency domain. Thus, corresponding GW templates are required as a priority to determine the detection band. For NS-NS and NS-WD binaries, the detection distance is at the order of $10^2$Mpc \cite{10.1093/mnras/staa2681,Hamilton_2022}, and we are expected to detect several or dozens of NS-NS merger signals in the future \cite{Zhu_2022}.

The Post-Newtonian approximation \cite{Kremer_2022,PhysRevD.81.064004,CC,DD,PhysRevD.87.121501} is widely used in the approximate solutions of the Einstein field equation, expanding it to higher order terms than Newtonian to make the solution more precise. It plays a significant role in the theoretical calculation of GWs from slow-moving objects in weak fields, which are in most cases in binary systems. The theoretical framework of the Post-Newtonian approximation approach to a binary system consists of two parts: binary dynamics and GW emissions \cite{PhysRevD.98.024039}. The post-Newtonian dynamical equations of this two-body problem generally take two forms \cite{A}: the Hamiltonian form and the Lagrangian form. In some situations, the Lagrangian form may give chaotic orbits, while the latter is always compatible with Keplerian-type parametric solutions that we are used to. These are also crucial parameters in calculating the GWF of binaries inspiralling along post-Newtonian accurate eccentric orbits \cite{PhysRevD.93.064031}. So our procedure is to first solve the dynamics equations, and then plug the orbit into the GW formula to obtain the waveform. 

GW signals generated by the inspiral and merger of binary stars are the main targets of future celestial GW detectors. And merger and ringdown phases of the BH-BH binary are always calculated by the numerical relativity \cite{PhysRevD.107.063029}, This paper aims to discuss the Post-Newtonian evolution of NS-NS and NS-WD binary systems and calculate their GWFs using the approximation. Notice that the mass range of NSs is generally between $1.35\thicksim2.1M_\odot$, and the radius is between $1.4\times10^{-5}\thicksim3.3\times10^{-5}R_\odot$, while WDs are much looser with a mass range generally between $0.4\thicksim1.2M_\odot$, and radius between $8\times10^{-3}\thicksim2\times10^{-2}R_\odot$, so more attention is paid on the NS-WD binaries when we investigate the effect of mass-transfer \cite{Krolik_2022,mass_trans,Chen_2022}. 

As for the outcome of the NS-NS merger, it varies depending on the angular momentum, mass ratio, equation of state, and other conditions. Usually, the total mass of an inner Milky Way NS-NS binary is within $2.5\thicksim2.88M_\odot$, and the mass ratio is between 0.77 and 0.91. When the total mass reaches $3.15\thicksim4.10M_\odot$, the binary may immediately collapse into a black hole with angular momentum density $a^*_{BH}=J/M$ around $0.7\thicksim0.8$ \cite{AAAA,PhysRevLett.129.181101}. The material of its accretion disk will eventually fall into the black hole along with the GW radiation and angular momentum loss in the ringdown phase. That is the situation we take into consideration at the end of evolution, whose GW can be calculated by black hole perturbation theory.

The theoretical analysis and numerical computation in this paper are arranged in the following sections. In Sec.\uppercase\expandafter{\romannumeral2}, we briefly review the standard routine to simplify the independent variables in Einstein field equation in the Transverse-Traceless (TT) gauge where the final metric perturbation $h_{\mu\nu}$ has only two independent variables $h_+(t)$ and $h_\times(t)$. In Sec.\uppercase\expandafter{\romannumeral3}, we give the evolution equation of the binary system and use the 2.5PN approximation of Hamiltonian form to calculate the trajectory and radiation power. In Sec.\uppercase\expandafter{\romannumeral4}, we propose the tidal disruption model \cite{SizeofFragments,universe8110576,PhysRevLett.129.121101}, and mass overflow model \cite{izzard_hall_tauris_tout_2011,10.1111/j.1365-2966.2006.11386.x}, where the GW of the mass-background (common envelope and accretion disc \cite{frank2002accretion}) is treated as a rotating non-spherically symmetric star\cite{Wong_2021,10.1093/mnras/stab3200}. Finally, waveforms concerning different branch parameters and initial mass ratios are discussed. All calculations are in natural units where $G=c=1$.

\section{TWO POLARIZATION MODES}

We begin with the Einstein field equation, where the metric $g_{\mu\nu}$ is a symmetric (0,2) tensor with generally 10 independent variables constrained by 10 correlated second-order nonlinear partial differential equations
\begin{equation}
G_{\mu\nu}(g_{\mu\nu})=\kappa T_{\mu\nu}.
\end{equation}
Due to the conservation of energy-momentum tensor, 4 constraints are eliminated by Bianchi identity 
\begin{equation}
\nabla^\mu G_{\mu\nu}=\nabla^\mu T_{\mu\nu}=0.
\end{equation}
The independence of coordinates also looses 4 constraints, leaving only two constraints with actual physical meaning. To obtain the specific metric with those two variables, supplemental constraints are required along with the Einstein field equation, which is the harmonic gauge and Transverse-Traceless (TT) gauge \cite{DanielG.Figueroa_2011} as
\begin{equation}
h^{TT}_{ij}=h^{TT}_{ji},\;\;\sum_ih^{TT}_{ii}=0,\;\;\sum_i\nabla_ih^{TT}_{ij}=0.
\end{equation}

Writing down the spacetime metric as $g_{\mu\nu}=\eta_{\mu\nu}+h_{\mu\nu}$, where $\eta_{\mu\nu}$ is the flat metric and $h_{\mu\nu}$ are small perturbations, the former reduces Einstein field equation to
\begin{gather}
\square\widetilde{h}_{\mu \nu}=-2\kappa T_{\mu \nu},\;\; \widetilde{h}_{\mu \nu} = h_{\mu \nu}-\frac{h}{2}\eta_{\mu \nu}.
\end{gather}
Choosing the GW transverse to z axis, the latter gives
\begin{equation}
h^{TT}_{ab}(t,z)=\begin{pmatrix}C_+ & C_\times \\
C_\times & -C_+\end{pmatrix}\cos\left[\omega\left(t-\frac{z}{c}\right)\right],
\end{equation}
where $a,b=1,2$, $z$ is the direction of propagation, and $C$ is the time independent strain of GW. Set $h_{ab}\equiv C_{ab}\cdot\cos[\omega(t-\frac{z}{c})]$, then we can get the metric
\begin{gather}
\begin{split}
\mathrm{d}s^2=&-c^2\mathrm{d}t^2+\mathrm{d}z^2+(1+h_+ )\mathrm{d}x^2\\
&+(1-h_+ )\mathrm{d}y^2+2h_\times  \mathrm{d}x\mathrm{d}y.
\end{split}
\end{gather}
Based on this metric, we can get the equation of $h_+(t)$ and $h_\times(t)$ by calculating the perturbation of metric tensor.

\section{BINARY EVOLUTION AND GWFS}

\subsection{2.5 PN Approximation}
To calculate the trajectory and GWF in the near-merger phase where the orbiting velocity is large, we need Post-Newtonian expansions for higher-order terms of velocity. Ignoring the spin of stars, here we adopt expansions up to 2.5PN \cite{Blanchet}, the metric can be written as
\begin{align}
\begin{split}	g_{00}=&-1+\frac{2}{c^2}V-\frac{2}{c^4}V^2+\frac{8}{c^6}\left( \hat{X}+V_iV_i+\frac{V^3}{6} \right)\\
& +\mathcal{O} \left( \frac{1}{c^{8}} \right),
\end{split}\\	
g_{0i}=&-\frac{4}{c^3}V_i-\frac{8}{c^5}\hat{R}_i+\mathcal{O} \left( \frac{1}{c^7} \right),\\
\begin{split}	g_{ij}=&\delta _{ij}\left[ 1+\frac{2}{c^2}V+\frac{2}{c^4}V^2\right] +\frac{4}{c^4}\hat{W}_{ij}+\mathcal{O} \left( \frac{1}{c^6} \right),
\end{split}\label{metric}
\end{align}
where $V$ and $V_i$ are the Newtonian potentials, $\hat{X}, \hat{R}_i$, and $\hat{W}_{ij}$ are the 2PN retarded potentials of gravity.

Usually, the pressure and temperature are negligible for regular compact binaries in weak fields \cite{Kremer_2022,PhysRevD.69.104021}. Here we only consider the dynamic equations of compact binaries due to mass distribution. Parameters are introduced as: the total mass of system $M=m_p+m_c$, the lost mass $\delta m_c=m^{(0)}_{c}-m_c$ of $m_c$, where $m^{(0)}_{c}$ is the initial mass of companion star and $m_c$ is the final mass. $\mu=m_pm_c/M$, $\nu=m_pm_c/M^2$, $q=m_c/m_p$ and in this paper we have $q\leqslant 1$. Let $\pmb{x}$ and $\pmb{v}=\mathrm{d}\pmb{x}/\mathrm{d}t$ be the relative coordinate vector and velocity vector, defining the unit vector $\pmb{n}=\pmb{x}/r$ where $r$ is the distance between two stars, and $\dot{r}\equiv \pmb{n}\cdot\pmb{v}$.

These post-Newtonian dynamics equations for the two-body problem can be written as Hamiltonian form or Lagrangian form \cite{A}. Here we use Hamiltonian to avoid chaotic results by Lagrangian form in harmonic gauge. Choosing the center of the primary star to be the origin, the 2.5PN equations of motion are obtained as \cite{Blanchet}

\begin{widetext}

\begin{equation}
    \frac{\mathrm{d}\pmb{v}}{\mathrm{d}t}=-\frac{GM}{r^2}[(1+\mathcal{A})\pmb{n}+\mathcal{B}\pmb{v}]+\mathcal{O}\left(\frac{1}{c^6}\right),
\end{equation}
where the coefficients of velocity direction $\mathcal{A}$ and radial direction $\mathcal{B}$ are:
\begin{align}
\begin{split}
    \mathcal{A}&=\frac{1}{c^2}\left\{-\frac{3\dot{r}^2\nu}{2}+v^2+3\nu v-\frac{GM}{r}(4+2\nu)\right\}\\
    &+\frac{1}{c^4}\left\{\frac{15\dot{r}^4\nu}{8}-\frac{45\dot{r}^4\nu}{8}-\frac{9\dot{r}^2\nu v^2}{2}+6\dot{r}^2\nu^2v^2+3\nu v^4-4\nu^2v^4\right.\\
    &\qquad\quad\left.+\frac{GM}{r}\left(-2\dot{r}^2-25\dot{r}^2\nu-2\dot{r}^2\nu^2-\frac{13\nu v^2}{2}+2\nu^2 v^2\right)+\frac{G^2M^2}{r^2}\left(9+\frac{87\nu}{4}\right)\right\}\\
    &+\frac{1}{c^5}\left\{-\frac{GM}{r}\frac{24\dot{r}\nu v^2}{5}-\frac{G^2M^2}{r^2}\frac{136\dot{r}\nu}{15}\right\},
\end{split}\\
\nonumber\\
\begin{split}
    \mathcal{B}&=\frac{1}{c^2}\bigg\{-4\dot{r}+2\dot{r}\nu\bigg\}\\
    &+\frac{1}{c^4}\left\{\frac{9\dot{r}^3\nu}{2}+3\dot{r}^3\nu^2-\frac{15\dot{r}\nu v^2}{2}-2\dot{r}\nu^2v^2+\frac{GM}{r}\left(2\dot{r}+\frac{41\dot{r}\nu}{2}+4\dot{r}\nu^2\right)\right\}\\
    &+\frac{1}{c^5}\left\{\frac{GM}{r}\frac{8\nu v^2}{5}+\frac{G^2M^2}{r^2}\frac{24\nu}{5}\right\}.
\end{split}
\vspace{0.1in}
\end{align}
Where the terms of 1PN and 2PN will bring the precession effect, and 2.5PN will bring the dissipation effect of gravitational radiation. Furthermore, by combining the Kepler orbital perturbation with the post-Newton correction, the energy of the binary system can be obtained

\begin{equation}
    E=-\frac{\mu c^2x}{2}\left[1+\left(-\frac{3}{4}-\frac{\nu}{12}\right)x^2+\left(-\frac{27}{8}+\frac{19\nu}{8}-\frac{\nu^2}{24}\right)x^3\right]+\mathcal{O}\left(\frac{1}{c^8}\right).
\end{equation}
And the evolution equation of orbit angular momentum is
\begin{equation}
    J=\frac{G\mu M}{c x^{1/2}}\left[1+\left(\frac{3}{2}+\frac{\nu}{6}\right)x+\left(\frac{27}{8}-\frac{19\nu}{8}+\frac{\nu^2}{24}\right)x^2\right]+\mathcal{O}\left(\frac{1}{c^6}\right).
\end{equation}
\vspace{0.1in}
\end{widetext}
which can be calculated with frequency-related parameter $x$ as
\begin{gather}
    x\equiv\left(\frac{GM\Omega}{c^3}\right)^{2/3}=\mathcal{O}\left(\frac{1}{c^2}\right).
\end{gather}
where $\Omega=\omega/2\pi$ is the angular frequency.

We can also use the definitions $E\equiv m_c v^2/2-GM/r$ and $J\equiv \mu\omega r^2$, where $\omega$ is the rotation velocity.  The energy and angular momentum calculated in this way will be different from the results of the Kepler orbit due to the presence of precession. For orbits with eccentricity, we calculate the angular velocity $\omega=\pmb{v}\cdot\pmb{n}/r$ by the tangential velocity around the orbit. The instantaneous orbit eccentricity is calculated by system energy and angular momentum
\begin{equation}
    e=\left(1+\frac{2EJ^2}{G^2m_p^2m_c}\right)^{1/2}.
\end{equation}

Through the above equations, the dynamic evolution process of the binary star system and the corresponding orbital parameters in the dissipative system with gravitational radiation can be described.

\subsection{GWs from Inspiral to Merger}
Due to the change of mass distribution, the metric perturbation $h_{\mu\nu}$ as a function of time on the flat space-time background $\eta_{\mu\nu}$. The two polarization GW can be represented by the unit polarization vector $\pmb{P}, \pmb{Q}$, which is orthogonal to the transverse direction $\pmb{N}$:
\begin{align}
    h_+&=\frac{1}{2}(P_iP_j-Q_iQ_j)H^{TT}_{ij},\\
    h_\times&=\frac{1}{2}(P_iQ_j+P_jQ_i)H^{TT}_{ij},
\end{align}
Based on that, we can get
\begin{equation}
    h_{(+,\times)}=\frac{2G\mu x}{c^2R}\sum_{p=0}^{+\infty}x^{p/2}H_{p/2(+,\times)}(\psi,c_i,s_i;\ln x),
\end{equation}
which is the expression of the gravitational waves strain in the (l, m)=(2, 2) mode in terms of the changing rate of the mass quadrupole moment and rotation frequency. And $R$ is the distance from the detector to the source, which does not affect the shape of the gravitational wave waveform and can be chosen at will. For example, that for the GW150914 is $R=1.768\times10^{20}R_\odot$. 

The gravitational wave phase $\psi$ is expressed as
\begin{gather}
    \psi=\phi-\frac{2GM_{\text{ADM}}\Omega}{c^3}\ln\left(\frac{\Omega}{\Omega_0}\right),\\
    M_{\text{ADM}}=M\left[1-\frac{\nu}{2}\gamma+\frac{\nu}{8}(7-\nu)\gamma^2+\mathcal{O}\left(\frac{1}{c^6}\right)\right],
\end{gather}
where $\gamma$ is the post-Newtonian parameter:
\begin{equation}
    \gamma=x\left[1+\left(1-\frac{\nu}{3}\right)x+\left(1-\frac{65\nu}{12}\right)x^2\right].
\end{equation}
Here $M_{\text{ADM}}$ is the ADM mass of binary and it is very important to include its relevant post-Newtonian contributions. $\Omega_0$ is the constant frequency that comes from the integration constant, for instance, to be the entry frequency of some detector. Define $i$ as the observation angle, let $si = \sin i, ci=\cos i$, and $\Delta=(m_p-m_c)/M$ for the plus polarization we have \cite{Blanchet}

\begin{widetext}
\begin{align}
H_0&=-(1+c_i^{2})\cos2\psi-\frac{1}{96}s_i^{2}(17+c_i^{2})\\
\nonumber\\
H_{1/2}&=-s_i\Delta\left\{\cos2\psi\left(\frac{5}{8}+\frac{1}{8}c_i^{2}\right)-\cos3\psi\left(\frac{9}{8}+\frac{9}{8}c_i^{2}\right)\right\}\\
\nonumber\\
H_1&=\cos2\psi\left\{\frac{19}{6}+\frac{3}{2}c_i^{2}-\frac{1}{3}c_i^{4}+\nu\left(-\frac{19}{6}+\frac{11}{6}c_i^{2}+c_i^{4}\right)\right\}-\cos4\psi\left\{\frac{4}{3}s_i^{2}\left(1+c_i^{2}\right)\left(1-3\nu\right)\right\}\\
\nonumber\\
\begin{split}
H_{3/2}&=s_i\Delta\cos\psi\left\{\frac{19}{64}+\frac{5}{16}c_i^{2}-\frac{1}{192}c_i^{4}+\nu\left(-\frac{49}{96}+\frac{1}{8}c_i^{2}+\frac{1}{96}c_i^{4}\right)\right\}+\cos2\psi\bigg\{-2\pi\left(1+c_i^{2}\right)\bigg\}\\
&+s_i\Delta\cos3\psi\left\{-\frac{657}{128}-\frac{45}{16}c_i^{2}+\frac{81}{128}c_i^{4}+\nu\left(\frac{225}{64}-\frac{9}{8}c_i^{2}-\frac{81}{64}c_i^4\right)\right\}+s_i\Delta\cos5\psi\left\{\frac{625}{384}s_i^2\left(1+c_i^2\right)\left(1-2\nu\right)\right\}
\end{split}\\
\nonumber\\
\begin{split}
H_2&=\pi s_i\Delta\cos\psi\left\{-\frac{5}{8}-\frac{1}{8}c_i^2\right\}\\
&+\cos2\psi\left\{\frac{11}{60}+\frac{33}{10}c_i^{2}+\frac{29}{24}c_i^{4}-\frac{1}{24}c_i^{6}+\nu\left(\frac{353}{36}-3c_i^{2}-\frac{251}{72}c_i^{4}+\frac{5}{24}c_i^{6}\right)\right.\left.+\nu^{2}\left(-\frac{49}{12}+\frac{9}{2}c_i^{2}-\frac{7}{24}c_i^{4}-\frac{5}{24}c_i^{6}\right)\right\}\\
&+\pi s_i\Delta\cos3\psi\left\{\frac{27}{8}\left(1+c_i^{2}\right)\right\}\\
&+\frac{2}{15}s_i^{2}\cos4\psi\left\{59+35c_i^{2}-8c_i^{4}-\frac{5}{3}\nu\left(131+59c_i^2-24c_i^4\right)+5\nu^{2}\left(21-3c_i^{2}-8c_i^{4}\right)\right\}\\
&+\cos6\psi\left\{-\frac{81}{40}s_i^{4}\left(1+c_i^{2}\right)\left(1-5\nu+5\nu^{2}\right)\right\}\\
&+s_i\Delta\sin\psi\left\{\frac{11}{40}+\frac{5\ln2}{4}+c_i^{2}\left(\frac{7}{40}+\frac{\ln2}{4}\right)\right\}+s_i\Delta\sin3\psi\left\{\left(-\frac{189}{40}+\frac{27}{4}\ln(\frac{3}{2})\right)\left(1+c_i^{2}\right)\right\},
\end{split}
\end{align}
\begin{align}
\begin{split}
H_{5/2}&=s_i\Delta \cos\psi\left\{\frac{1771}{5120}-\frac{1667}{5120}c_i^2+\frac{217}{9216}c_i^4-\frac{1}{9216}c_i^6+\nu\left(\frac{681}{256}+\frac{13}{768}c_i^2-\frac{35}{768}c_i^4+\frac{1}{2304}c_i^6\right)\right.\\
&\qquad\qquad\qquad\left.+\nu^2\left(-\frac{3451}{9216}+\frac{673}{3072}c_i^2-\frac{5}{9216}c_i^4-\frac{1}{3072}c_i^6\right)\right\}\\
&+\pi\cos2\psi\left\{\frac{19}{3}+3c_i^2-\frac{2}{3}c_i^4+\nu\left(-\frac{16}{3}+\frac{14}{3}c_i^2+2c_i^4\right)\right\}\\
&+s_i\Delta\cos3\psi\left\{\frac{3537}{1024}-\frac{22977}{5120}c_i^2-\frac{15309}{5120}c_i^4+\frac{729}{5120}c_i^6+\nu\left(-\frac{23829}{1280}+\frac{5529}{1280}c_i^2+\frac{7749}{1280}c_i^4-\frac{729}{1280}c_i^6\right)\right.\\
&\qquad\qquad\qquad\left.+\nu^2\left(\frac{29127}{5120}-\frac{27267}{5120}c_i^2-\frac{1647}{5120}c_i^4+\frac{2187}{5120}c_i^6\right)\right\}\\
&+\cos4\psi\left\{-\frac{16\pi}{3}\left(1+c^2_i\right)s_i^2\left(1-3\nu\right)\right\}\\
&+s_i\Delta\cos5\psi\left\{-\frac{108125}{9216}+\frac{40625}{9216}c_i^2+\frac{83125}{9216}c_i^4-\frac{15625}{9216}c_i^6+\nu\left(\frac{8125}{256}-\frac{40625}{2304}c_i^2-\frac{48125}{2304}c_i^4+\frac{15625}{2304}c_i^6\right)\right.\\
&\qquad\qquad\qquad\left.+\nu^2\left(-\frac{119375}{9216}+\frac{40625}{3072}c_i^2+\frac{44375}{9216}c_i^4-\frac{15625}{3072}c_i^6\right)\right\}\\
&+\Delta\cos7\psi\left\{\frac{117649}{46080}s_i^5\left(1+c_i^2\right)\left(1-4\nu+3\nu^2\right)\right\}+\sin2\psi\left\{-\frac{9}{5}+\frac{14}{5}c_i^2+\frac{7}{5}c_i^4+\nu\left(32+\frac{56}{5}c_i^2-\frac{28}{5}c_i^4\right)\right\}\\
&+s_i^2\left(1+c_i^2\right)\sin4\psi\left\{\frac{56}{5}-\frac{32\ln2}{3}+\nu\left(-\frac{1193}{30}+32\ln2\right)\right\}.    
\end{split}
\end{align}

\end{widetext}

\subsection{GWs in Ringdown Phase}
In order to obtain the complete GWF, we consider binaries merge to form a black hole. The evolutionary process of binaries can be cut into three phases inspiral-merger-ringdown by jump functions
\begin{equation}
h=h^{IM}\Theta(t_0-t)+h^{RD}\Theta(t-t_0),
\end{equation}
where $t_0$ is the time of merger, $\Theta$ is jump function of $t$. if $t_0-t>0$, $\Theta(t_0-t)=1$, for the else $\Theta(t_0-t)=0$. The GW of ringdown $h^{RD}$ is calculated by black hole perturbation theory. In late ringdown trailing phase, we have \cite{PhysRevLett.129.111102,PhysRevD.100.044018,PhysRevLett.123.161101}
\begin{equation}
    h^{lm}=\frac{2GM\nu x}{Rc^2}\sqrt{\frac{16\pi}{5}}H^{lm}e^{-\pi \omega t/\mathcal{Q}}\cos(\omega t).
\end{equation}
When it is far enough the above equation can be approximated as
\begin{gather}
\mathcal{Q}=2\bigg(1-a_*\bigg)^{-\frac{9}{20}}=2\left(1-\frac{c\pmb{J}}{GM^2}\right)^{-\frac{9}{20}},\\
a_*=c\pmb{J}/GM^2=\pmb{J}/M^2,
\end{gather}
$a_*$ is the spin rate of black hole \cite{PhysRevD.40.3194}. The angular momentum $\pmb{J}=\pmb{r}_i\cdot\pmb{p}_i$, where $\pmb{r}_i$, $\pmb{p}_i$ is the winding radii and momentum of binaries. The usual compact binary systems in the Milky Way, like GS2000+25 and LMC X-3 systems, the values of their dimensionless spin parameters are both $a_*=0.03$ \cite{10.1046/j.1365-8711.1999.02482.x}. 

\begin{equation}
\begin{split}
    H^{22}&=1+x\left(-\frac{107}{42}+\frac{55}{42}\nu\right)+2\pi x^{3/2}\\
    &+x^2\left(-\frac{2173}{1512}-\frac{1069}{216}\nu+\frac{2047}{1512}\nu^2\right)\\
    &+x^{5/2}\left(-\frac{107\pi}{21}-24i\nu+\frac{34\pi}{21}\nu\right)+\mathcal{O}\left(\frac{1}{c^6}\right).
\end{split}
\end{equation}

Since mass-transfer mainly happens in inspiral and merger, our discussion on the GWF and binary evolution correction are focused on that period. To make the whole process more complete, we extend our GWF to ringdown phase $h^{RD}$, which also varies as the mass-transfer correction is applied to the common envelope evolution mode.

\subsection{Binary System Evolution Equations}

Based on the results above, we set the gravitational potential $E$, GW frequency $f$, and rotation period $T$ as functions of time. Plugging into formulas we obtain the average radiation power $\mathcal{\bar{F}}$, then we get the average orbital parameters in Keplerian orbit with Newton approximation as
\begin{align}
\frac{\mathrm{d}a}{\mathrm{d}t}&=-\frac{64}{5}\frac{\mu M^2}{a^3(1-e^2)^{\frac{7}{2}}}\left(1+\frac{73}{24}e^2+\frac{37}{96}e^4\right),\\
\frac{\mathrm{d}e}{\mathrm{d}t}&=-\frac{304}{15}\frac{e\mu M^2}{a^4(1-e^2)^{\frac{5}{2}}}\left(1+\frac{121}{304}e^2\right),
\end{align}
\begin{align}
\frac{\mathrm{d}T}{\mathrm{d}t}&=-\frac{96}{5}\frac{T\mu M^2}{a^4(1-e^2)^{\frac{7}{2}}}\left(1+\frac{73}{24}e^2+\frac{37}{96}e^4\right).
\end{align}
These orbital parameters can help us to analyze the change of the trajectory, but the final motion equation is still given by the 2.5PN correction. Here we set the initial values as
\begin{gather}
m_p=2 [M],\;\;m_c=1.4 [M],\nonumber\\
r_0=a_0(1+e)=50 [M]^{-1},\;\;R_{NS}=7[M]^{-1},\nonumber\\
e=0.25,\;0.30,\;0.35.\nonumber
\end{gather}

\begin{figure}[ht!]
    \centering
    \includegraphics[width=1\linewidth]{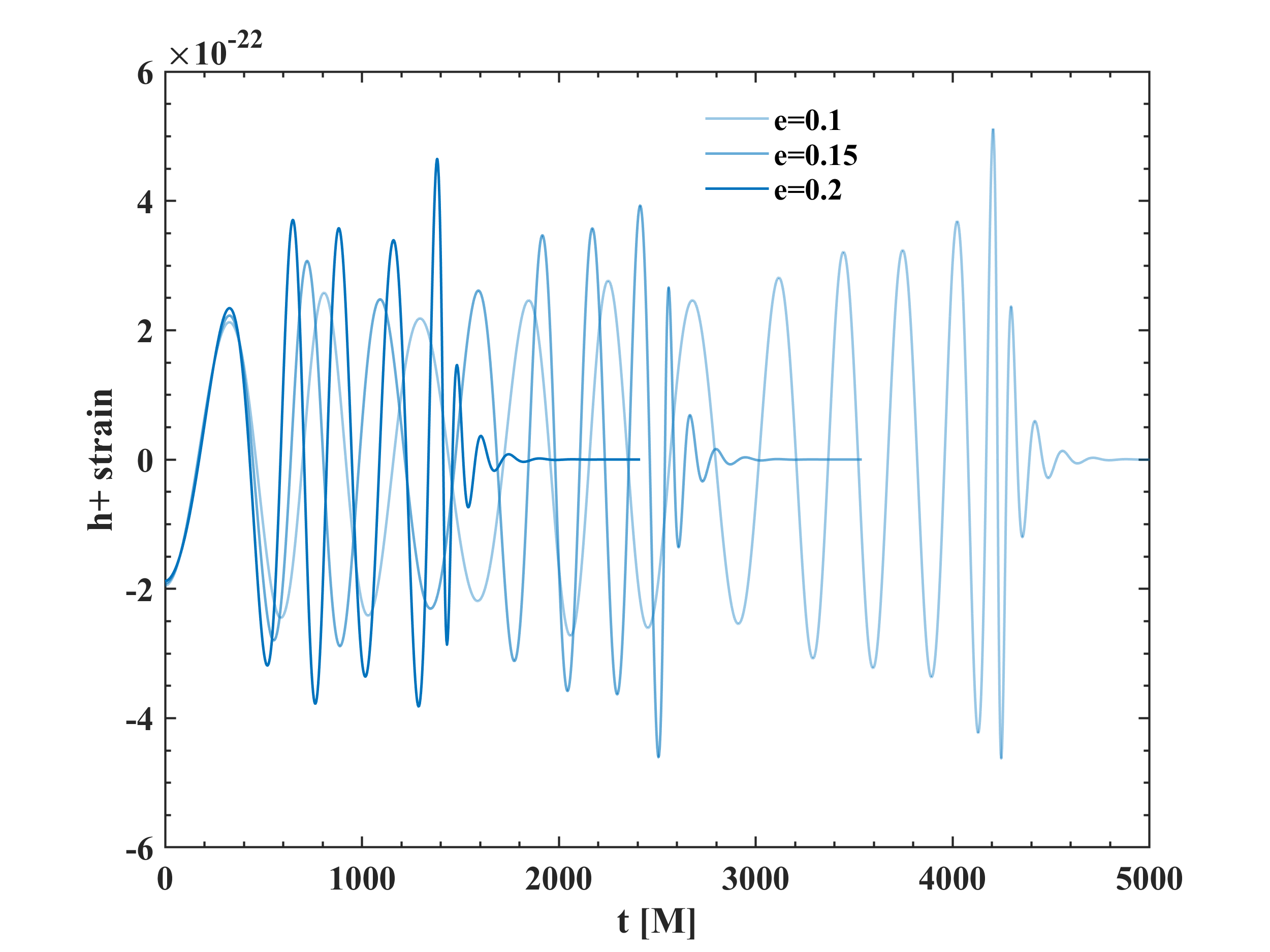}
    \caption{\small This figure shows the terminal GWFs of NS-NS systems with different initial eccentricities, which shows the negative correlation between the eccentricity and orbiting time.} 
	\label{F1}
\end{figure}

The 4th-order Rounge-Kutta method is used to solve those 2.5PN differential equations numerically. The GWFs are shown as follows by plugging data into formulas. A comparative figure of different initial eccentricities of NS-NS binaries without mass-transfer is given in FIG.\ref{F1}. 

 To better reflect the evolution of the binary motion equation and the corresponding GWF at this initial distance, we will take $e=0.3$ as the initial eccentricity in the subsequent calculation.

\section{MASS TRANSFER CORRECTION}

Due to the violent fraction and distortion during the transfer, much heat may accumulate inside the accretion disk, triggering a nuclear reaction and causing supernova-like detonation \cite{radice2020dynamics}. However, the WD nuclear burst is not inevitable. It requires some ignition conditions, such as temperature, accretion rate, initial mass of WD, and so on. When the heat dissipation is good \cite{Fernández_2013,Dohi__2021}, the mass of the WD is close to the Chandrasekhar mass or reaches central densities sufficiently high ($\rho_0 \gtrsim 10^{9.7}\thicksim10^{10} \text{g/cm}^3$), the WD may continue to collapse due to accretion and self-gravitation as "accretion-induced collapse" (AIC), and instead of supernova burst\cite{PhysRevD.81.044012}.

\begin{figure}[ht!]
    \centering
    \includegraphics[width=0.85\linewidth]{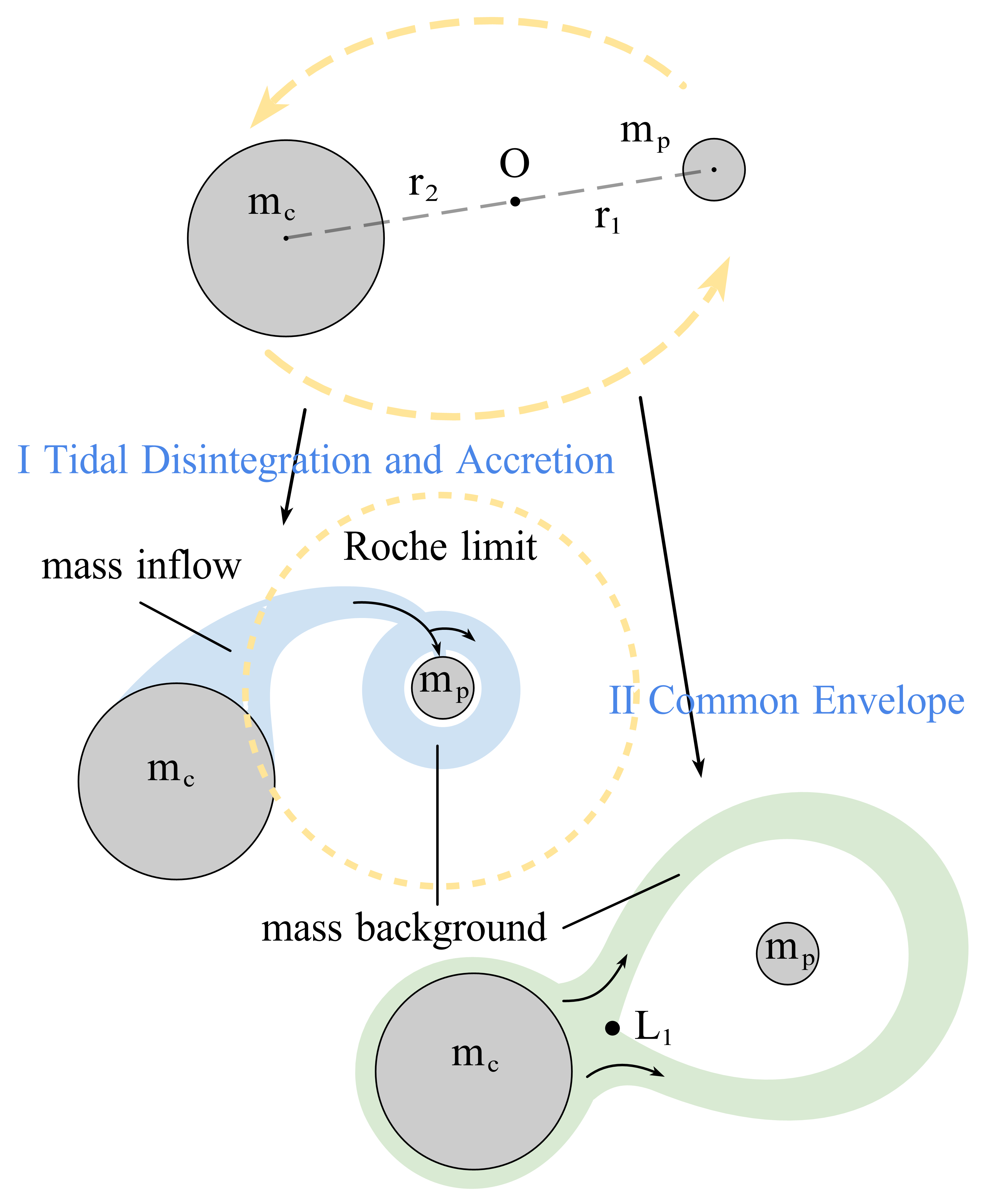}
    \caption{\small At the top of this figure is the compact binaries in inspiral, where the primary star $m_p$ is more compact than the companion star $m_c$. Below are two mass transfer models, the left one is the case with tidal disruption branches into and direct accretion flow, and the right one is the case of the common envelope. When calculating GWFs, the common envelope is regarded as an overall mass background.}
	\label{mass_transfer}
\end{figure}

In this paper we considered the case of stable mass transfer, which restricts our discussion to the WDs whose thermonuclear composition is not explosive, so all burning processes are quiescent enough to be negligible by dynamical means. Second, the state of the stars may alter in inspiral, especially when their mass changes. That’s to say, their volume may take a spontaneous expansion or contraction, passing through some critical boundary like the Roche lobe, resulting in discontinuous transfer. Again this intrinsic property is determined by stars’ nuclear structure, and we restrict to sufficient stable ones that show no such a behavior. Third, the non-gravity couplings, such as the magnet field of the NS and emitted light pressure, are assumed to be ignorable, so the whole transfer is dominated by gravity \cite{Belloni2022}.

\subsection{Mass Transfer Models}

We establish a model where there is a stable and conservative mass-loss flow between two stars and propose the following two models of mass-transfer: the mass overflow model and the tidal disruption model. For a typical star consisting of a highly dense core and a wrapping atmosphere, the mass transferring process can be summarized as Fig.\ref{mass_transfer} In real cases, like a WD, the process may be far more complicated than driven only by gravity as we considered here. The main material components considered include compact binary stars, the incompressible fluid for mass transfer, and the accretion disc.

The transferring process is constrained by the conservation of energy and mass. Driven only by gravity, we obtain the Euler equation of incompressible fluid
\begin{equation}
\frac{\partial}{\partial t}\left(\begin{gathered}
     \rho  \\
     \rho v_x\\
     \rho v_y\\
     \rho \epsilon
\end{gathered}\right)+\nabla\left(\begin{gathered}
     \rho v_x \\
     \rho v_x^2+P \\ 
     \rho v_xv_y\\ 
     (\rho \epsilon+P)v_x 
     \end{gathered}\;\;
     \begin{gathered}
     \rho v_y\\
     \rho v_yv_x\\
     \rho v_y^2+P\\
     (\rho \epsilon+P)v_y
\end{gathered}\right)=0.
\end{equation}
where $\epsilon$ is the energy per unit volume. Along with the two constrained equations, we can solve them to gain the mass-transfer speed of the common envelope and accretion process.  The mass transfer influence on the binary system mainly contains the following sections.

\subsection{Mass Overflow of the Atmosphere}

In the case of the common envelope evolution, we need to calculate the mass that flows from $m_c$ to $m_p$, and that becomes the mass background. Here we introduce two parameters to describe the stability of mass-transfer in NS-WD systems
\begin{equation}
\zeta_P=\partial\ln r_p/\partial\ln \dot{M},\;\;\zeta_L=\partial\ln r_L/\partial\ln \dot{M},
\end{equation} 
which are associated with the adiabatic mass-radius $r_p$ and the NS Roche lobe radius $r_L$ \cite{1989The,1983ApJ...268..368E}.
When $\zeta_P-\zeta_L>0$ we think it is a stable mass-transfer process. Actually, three effects are competing for the transferring switch. Newtonian angular momentum stretches the separation between the two stars as mass ratio $q$ decreases while the gravitational radiation is contradictory. Also, the Roche lobe itself has to do with mass ratio, so this is just a calculation-convenient condition. 

As the binary radiates its angular momentum, the exterior of the donor's atmosphere may exceed the border of the shrinking Roche lobe,  causing an overflow of its mass \cite{1975Gas,10.1093/mnras/stad1862}. The total mass of the overflow can be expressed by the majority amount passing through the $L1$ Lagrangian times a coefficient $Q_\rho$. 
\begin{equation}
    Q_\rho=\frac{2\pi}{\omega^2\sqrt{A(A-1)}}c_T^2,\;\;c_T=\frac{k_B}{\bar{m}_g}T_0.
\end{equation}Here $c_T$ is the isothermal sound speed at $L1$, $\bar{m}_g$ is the mean mass of a gas particle, $T_0$ is its temperature and $k_B$ is the Boltzmann constant. Parameter $A$ is defined as
\begin{align}
A=4+\frac{104/25}{24/25+q^{1/3}+q^{-1/3}}.
\end{align}

The overflow pervades filling binary Roche lobe and eventually forms a common envelope becoming a homogeneous mass background
\begin{equation}
\dot{m_{c}}=-\frac{2\pi}{\omega^2\sqrt{A(A-1)}}c_T^2v_0\rho_0,
\end{equation}
where $v_0$ is the stream speed passing through $L1$. The scale of $\dot m_c$ has to do with a specific case and in the following calculation we choose $c_T^2v_0= 1.28\times 10^{-14}c^3$.

\subsection{Tidal Disruption of the Core}
As the orbit shrinks to a smaller size,  the core experiencing growing tidal force starts to disintegrate. The debris forms an interstellar stream flowing towards the primary star. Before finally accreting on the primary star surface, the interstellar flow needs to first lose its angular momentum in the taken by friction. For a  priorly steady disc, this process can be equivalently viewed as some interstellar flow branches into the while some remain direct accretion flow.

Considering a straight in-compressible flow of mass-loss from $m_c$ star to $m_p$ star in the Roche limit of the more massive star. This mainly describes cases where the companion star is disrupted very apace or the equivalent current where the mass loss dramatically collides with the intermediate debris and shows as a macroscopic transport. In order to represent the strength of the mass-transfer, we introduced the Euler equations for incompressible fluids \cite{Euler,Euler1}.

\begin{equation}
\dot m_{c}=\frac{\mathrm{d}m_c}{\mathrm{d}v}\frac{\mathrm{d}v}{\mathrm{d}t}+\frac{\mathrm{d}m_c}{\mathrm{d}r}\frac{\mathrm{d}r}{\mathrm{d}t},
\end{equation}
\begin{equation}
\rho\frac{\partial}{\partial t}\begin{pmatrix}
      v_x\\
      v_y
\end{pmatrix}+\begin{pmatrix}
     -\rho g_x+\partial_x P &0\\ 
     0&-\rho g_y+\partial_y P
\end{pmatrix}=0,
\end{equation}where $P$ is the pressure of the mass flow, which will be calculated from the equation of state of the stars in the following sections, and $g_x$ and $g_y$ are the components of gravitational acceleration of the tidal gravity 
\begin{equation}
\pmb{g}=\frac{m_p(r-R_c)^2-m_cR_c^2}{r^2R_c^2-2rR_c^3+R_c^4}\pmb n,
\end{equation}
where $R_c$ is the radius of companion star. 

\subsection{The Accretion Disc}
In order to obtain the analytical mass transfer model, we assume that there is a stable rotationally symmetric accretion disk around the NS \cite{frank2002accretion,Franchini_2021}. Due to this rotational symmetry in the $x-y$ plane, it has no GW emission in the perpendicular direction. Therefore, we only consider the influence of the accretion disk (AD) on the angular momentum. In our model, the AD  compressed to the $x-y$ plane, the surface density and radial velocity can be expressed as 
\begin{align}
\rho_{disc} &=A_{\rho}\alpha^{-4/5} \dot m_{d\rightarrow p}^{7/10}m_p^{1/4} R_{AD}^{-{3/4}} f^{14/5},
\\
v_R&=A_{v}\alpha^{4/5} \dot m_{d\rightarrow p}^{3/10}m_p^{-{1/4}} R_{AD}^{-{1/4}} f^{-{14/5}},\\
f&=[1-({R_{NS}}/{R_{AD}})^{1/2}]^{1/4},
\end{align}
where $\dot m_{d\rightarrow p}$ is the mass change rate of neutron star caused by AD, $R_{AD}$ is the radius of disc and $R_{p}$ is the radius of the primary star. All aspects of the viscosity mechanism have been packed into the prescription parameter $\alpha \lesssim 1$, and some \cite{10.1111/j.1365-2966.2007.11556.x} indicates $\alpha\thicksim 0.1$ in ADs in cataclysmic variables.

By means of unit conversion, we obtain the constant coefficients in the above formulas are 
\begin{equation}
    A_{\rho}=87.2645,\;A_v=0.0016089.\nonumber
\end{equation}
 We can obtain the total mass of the AD through the mass transfer model, so the final variable $\dot{m}_{d\rightarrow p}$ in the stable accretion disk can be obtained by normalization.
\begin{equation}
    \dot{m}_{d\rightarrow p}=\left(\frac{m_{AD}A_\rho^{-1}\alpha^{4/5}}{\int\mathrm{d}\theta\int m_p^{1/4}R_{AD}^{-{3/4}} f^{14/5}\mathrm{d}R_{AD}}\right)^{10/7}.
\end{equation}
The tangential velocity of the accretion disk is determined by the stable circular orbit of the radius, which means that $v_\tau=\sqrt{Gm_p/R_{AD}}$.

\begin{figure}[ht!]
    \centering
    \subfigure{\includegraphics[width=0.51\linewidth]{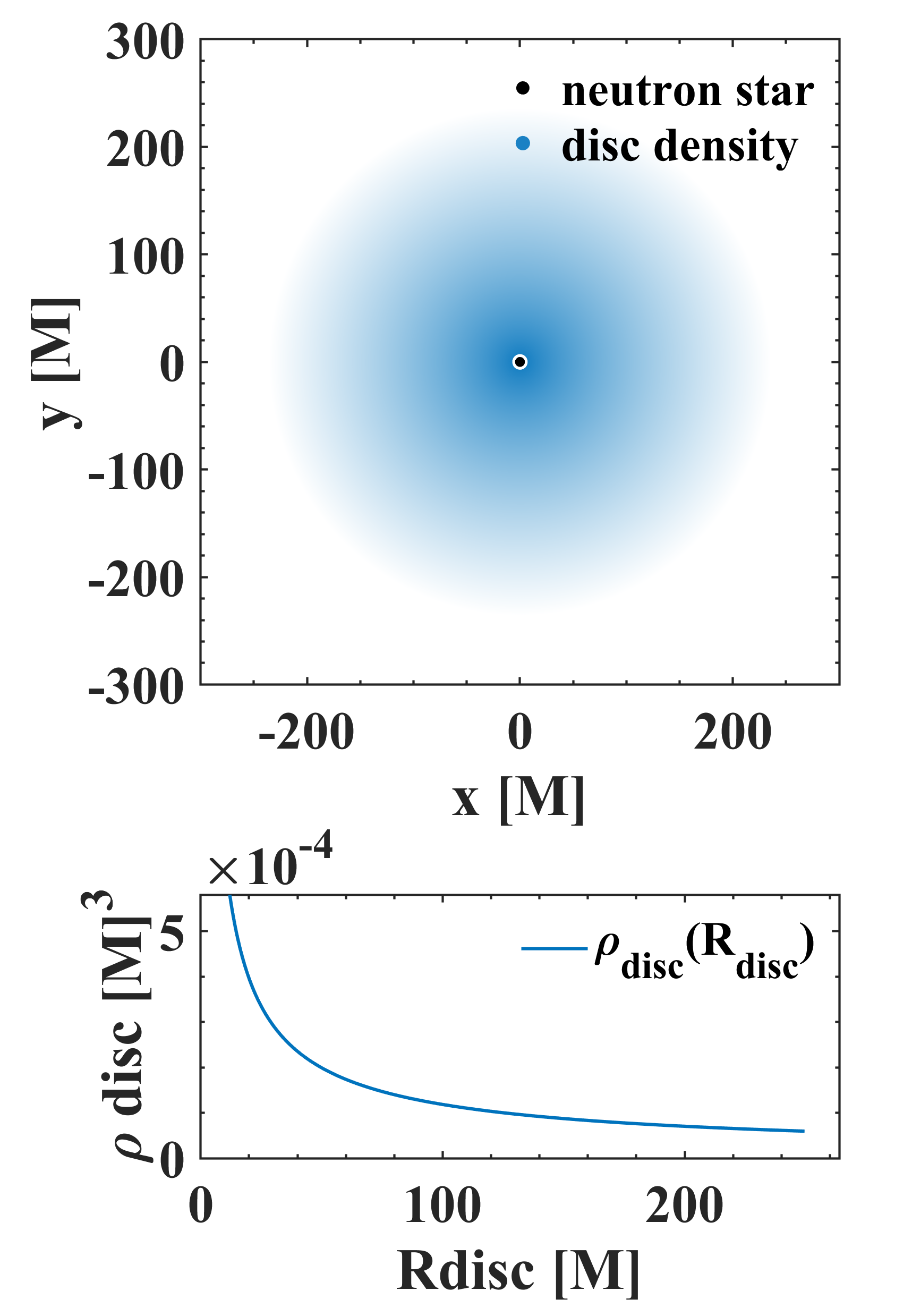}}
    \hspace{-0.2in}
    \subfigure{\includegraphics[width=0.51\linewidth]{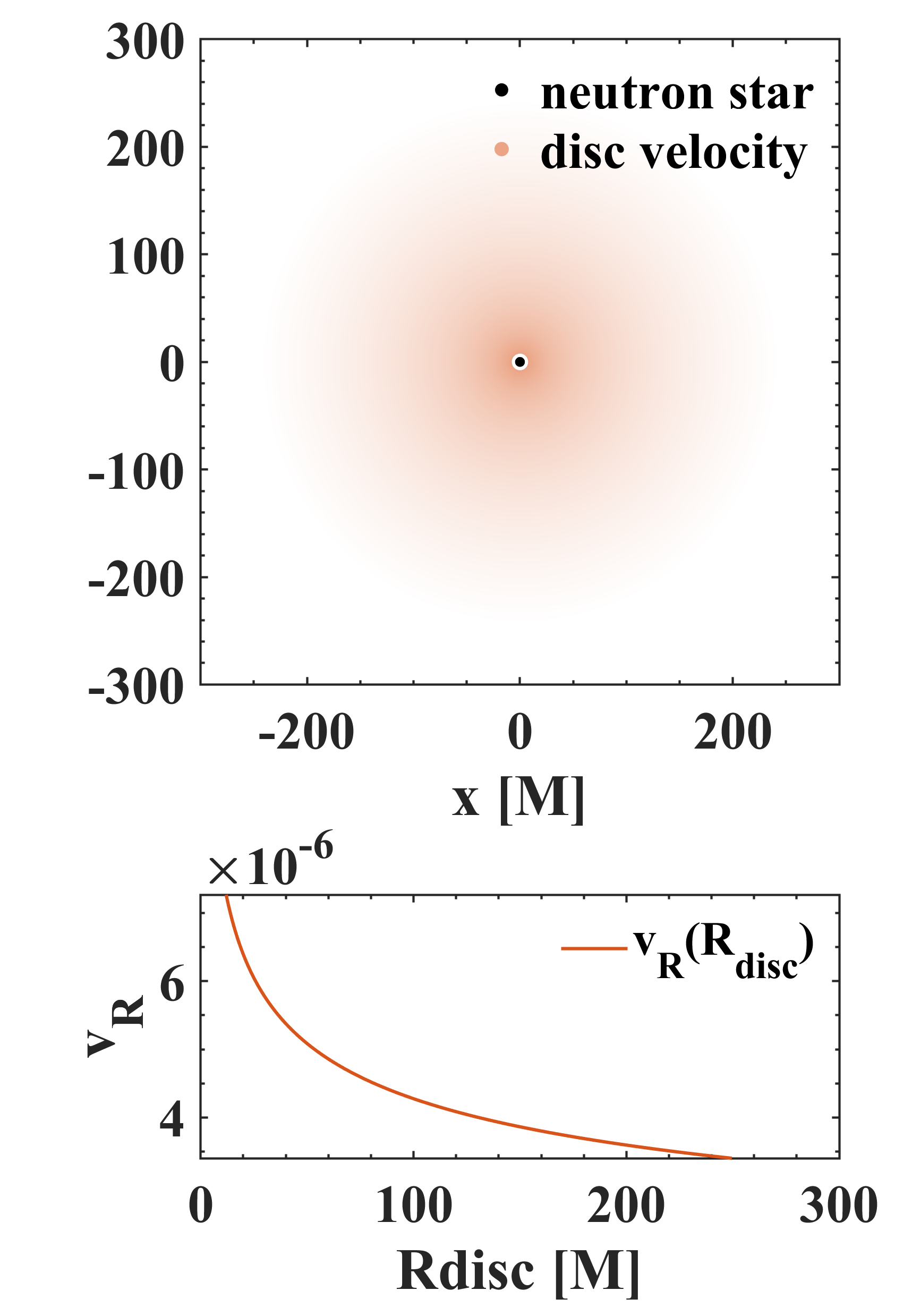}}
    \vspace{-0.1in}
    \caption{\small he left figure shows the surface density of the, which decays quickly as the radius gets larger. Thus the margin contribution is negligible, and we only plot 0.1 times the maximum density; The right figure shows the radial drift velocit indicating accretion, which also decays very quickly along the radius. Since the rotation of the disc is assumed to be Keplerian, the azimuthal velocity is trivial and not plotted.}
	\label{disc}
\end{figure}

Through the above equation, we obtain the density distribution and radial velocity distribution of the accretion disk as shown in FIG.\ref{disc}

 Then the density of the accretion disk is integrated into the whole space as
\begin{widetext}
\vspace{0.1in}
\begin{equation}
J_{AD}=\alpha^{-4/5}\dot m_{d\rightarrow p}^{7/10}m_p^{1/4}\int^{2\pi}_{0}\mathrm{d}\theta\int^{R_{o}}_{R_{i}}\mathrm{d}R_{AD}\left\{A_{\rho}R_{AD}^{1/4}\left[1-\left(\frac{R_{NS}}{R_{AD}}\right)^{1/2}\right]^{14/12}\sqrt{Gm_p/R_{AD}}\right\},
\end{equation}
where $R_o$ is the out radius of disc, $R_i=6m_p$ is the internal Inner Stable Circular Orbit (ISCO) radius. We know that $R_{p}\ll R_{AD}$, therefore, we expand the equation at $R = 0$, and obtain the density integral result under the first-order approximation as 
\begin{equation}
    J_{AD}=2\pi \alpha^{-4/5}\dot m_{d\rightarrow p}^{7/10}m_p^{1/4}\sqrt{Gm_{p}}\left(\frac{4}{3}R_{AD}^{3/4}-\frac{14}{5}R_{NS}^{1/2}R_{AD}^{1/4}\right)\Big|^{R_{o}}_{R_{i}}.
\end{equation}
\vspace{0.1in}
\end{widetext}
Through the conservation of the total angular momentum of the system, the angular momentum of the rotating orbit of the binary star system $J_L$ can be obtained as
\begin{equation}
    J_{L}\equiv\mu\omega r^2=J-\int \frac{J_{AD}}{m_{AD}}(\dot{m}_{AD}-\dot{m_{d\rightarrow p}})\mathrm{d}t.
\end{equation}
Among them, $J$ is the orbital angular momentum without accretion disk correction, which is given by the 2.5PN correction, including the total angular momentum loss caused by gravitational radiation. The $m_p$ term is the part of the accretion disk angular momentum converted into the spin angular momentum of the NS. Thus, the modified differential equation of tangential velocity ( or angular velocity ) can be obtained.

There will be dynamical friction \cite{10.1093/mnras/stu562,Adhikari_2016} when the $m_c$ star enters the common envelope that winds inside the primary star Roche lobe. This friction will obstruct the motion of $m_c$ star and cause kinetic energy loss, making the binary merge more rapidly. In our case, the frictional force is much weaker than the gravitational force and is negligible in the calculation.

\subsection{The Analytical Expression of Mass Transfer}
Through the analysis of the above sections, we combine mass overflow of the atmosphere with tidal disruption of the core, and get the final mass loss rate formula of WD is defined as $\dot m_c\equiv\dot m_{c1}+\dot m_{c2}$, and the specific analytical expression is
\begin{equation}
\dot m_c=-Q_\rho v_0\rho-\left(\frac{\partial V}{\partial v}\Big|\rho \pmb{g}-\nabla P\Big|+\rho\frac{\partial V}{\partial r}\frac{\partial r}{\partial t}\right). 
\end{equation}
Using tidal gravity as a mass transfer function, we then obtain the mass change in the system at any time interval
\begin{align}
m_p'=&m_p+(\kappa\dot{m}_c+\dot{m}_{d\rightarrow p})\mathrm{d}t,\\
m_c'=&m_c-\dot{m}_c\mathrm{d}t, 
\end{align}
where $0\leqslant \kappa\leqslant1$ is the branch parameter\cite{10.1093/mnras/stw1410}
\begin{equation}
  \kappa\equiv\partial \ln\dot{m}_c/\partial \ln r=|(\delta m_p-\delta m_{d\rightarrow p})/\delta m_c|.\\ 
\end{equation}
Here $\delta m_c$ is the mass-loss of the companion star and $\delta m_p$ is the primary star's gained mass. It can also be written as 
\begin{equation}
\begin{split}
\kappa(\eta_w,\gamma_{a})=&\frac{1}{2}\Big[1-\eta_w\gamma_{a}+(-5\gamma_{a}^2+10\eta_w\gamma_{a}^2+\eta_w^2\gamma_{a}^2\\
&\left.+6\gamma_{a})^{1/2}\right]/(-\gamma_{a}+2\eta_w\gamma_{a}+1),
\end{split}
\end{equation}
where $\gamma_a=5/3, 4/3$ is an adiabatic coefficient, and $\eta_w$  is the winding efficiency parameter, which depends on factors such as the mass ratio, Keplerian orbital velocity, and the asymptotic wind velocity. For $\eta_w\thicksim1$ this is a factor of $\vartheta<1$ times smaller than the local dynamical time-scale, where $\vartheta$ is a function of mass ratio $q$.

We know that for a $(1.4+0.1)M_\odot$ NS–WD system, $\kappa\gtrsim 0.5$ of the WD mass-loss can be captured by its companion NS, while for $(1.4+1.25)M_\odot$ NS–WD system, the branch parameter $\kappa$ drops steeply to $5.4\times 10^{-4}$ \cite{10.1093/mnras/stab626}. The relation between $\kappa$ and $q$ can be written in a more explicit form. Expanding  $\eta_w$ to the linear term of $q$, we can switch the above function into $\kappa(q,\gamma)$. In the following calculation, we set $\gamma=5/3$. 

Then, getting the evolution of inspiral semi-major axis and GW frequency with mass-transfer correction are \cite{10.1093/mnras/stab626}
\begin{align}
\frac{\mathrm{d}a}{\mathrm{d}t}&=-\frac{64\mu M^2a^{-3}}{5(1-e^2)^{\frac{7}{2}}}\left(1+\frac{73}{24}e^2+\frac{37}{96}e^4\right)-2aC\frac{\dot{m_c}}{m_c},\\
\frac{\mathrm{d}f}{\mathrm{d}t}&=\frac{96}{5}\pi^{\frac{8}{3}}m_pm_cM^{-\frac{1}{3}}+3fC^{-1}\frac{\dot{m_c}}{m_c},\\
C&=\left[1-\kappa q-(1-\kappa)\frac{q}{q+1}-\kappa\sqrt{1+q}\right]^{-1}.
\end{align}

After adding the mass-transfer correction, we calculated the GWFs of the NS-NS systems and the NS-WD systems.

\section{Numerical Results}

\subsection{NS-NS Systems}
The NS-NS binary system is a typical compact binary system. The masses of most NSs lie in $1.4\thicksim2.1 M_{\odot}$ and their densities are all close to each other. Therefore, the tidal disruption can only happen when the relative distance is very close. Here we calculate a case of extreme mass ratio. The thermonuclear reaction, electromagnetic effect, and the change of the equation of the state of NS are not considered.

\begin{figure}[ht!]
    \centering
    \subfigure{\includegraphics[width=1\linewidth]{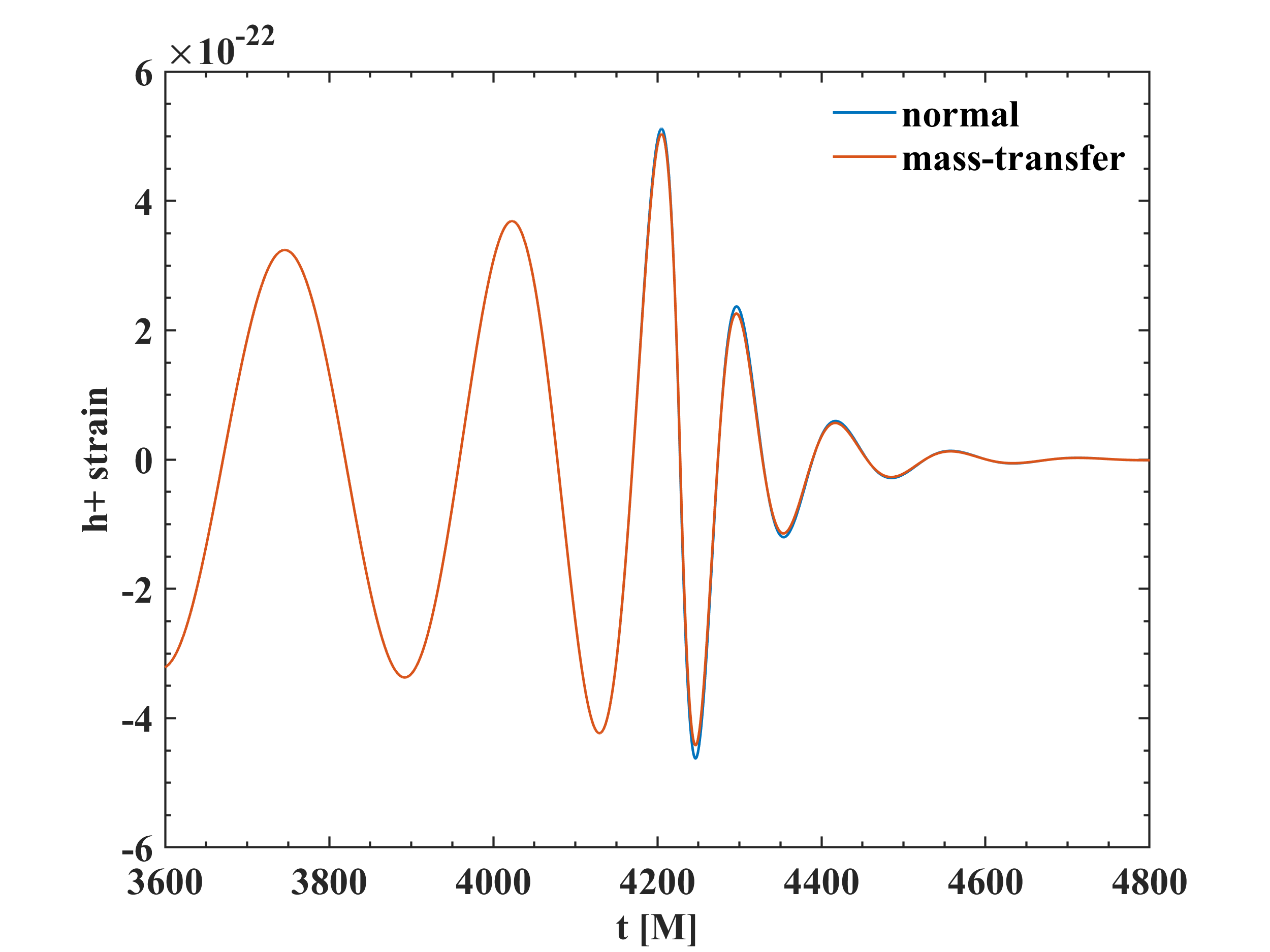}}\\
    \vspace{-4mm}
    \subfigure{\includegraphics[width=0.5\linewidth]{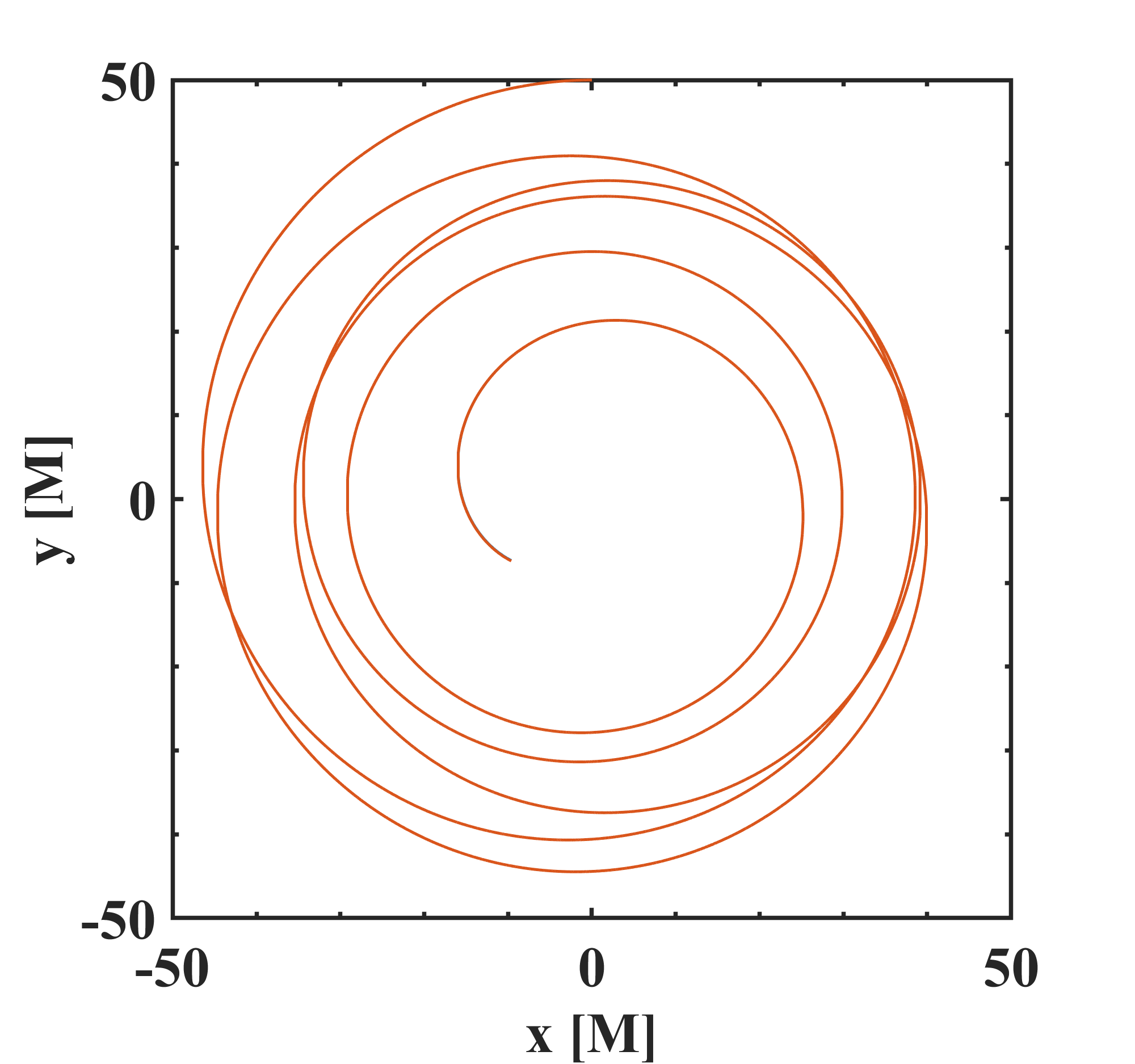}}
    \hspace{-0.1in}
    \subfigure{\includegraphics[width=0.5\linewidth]{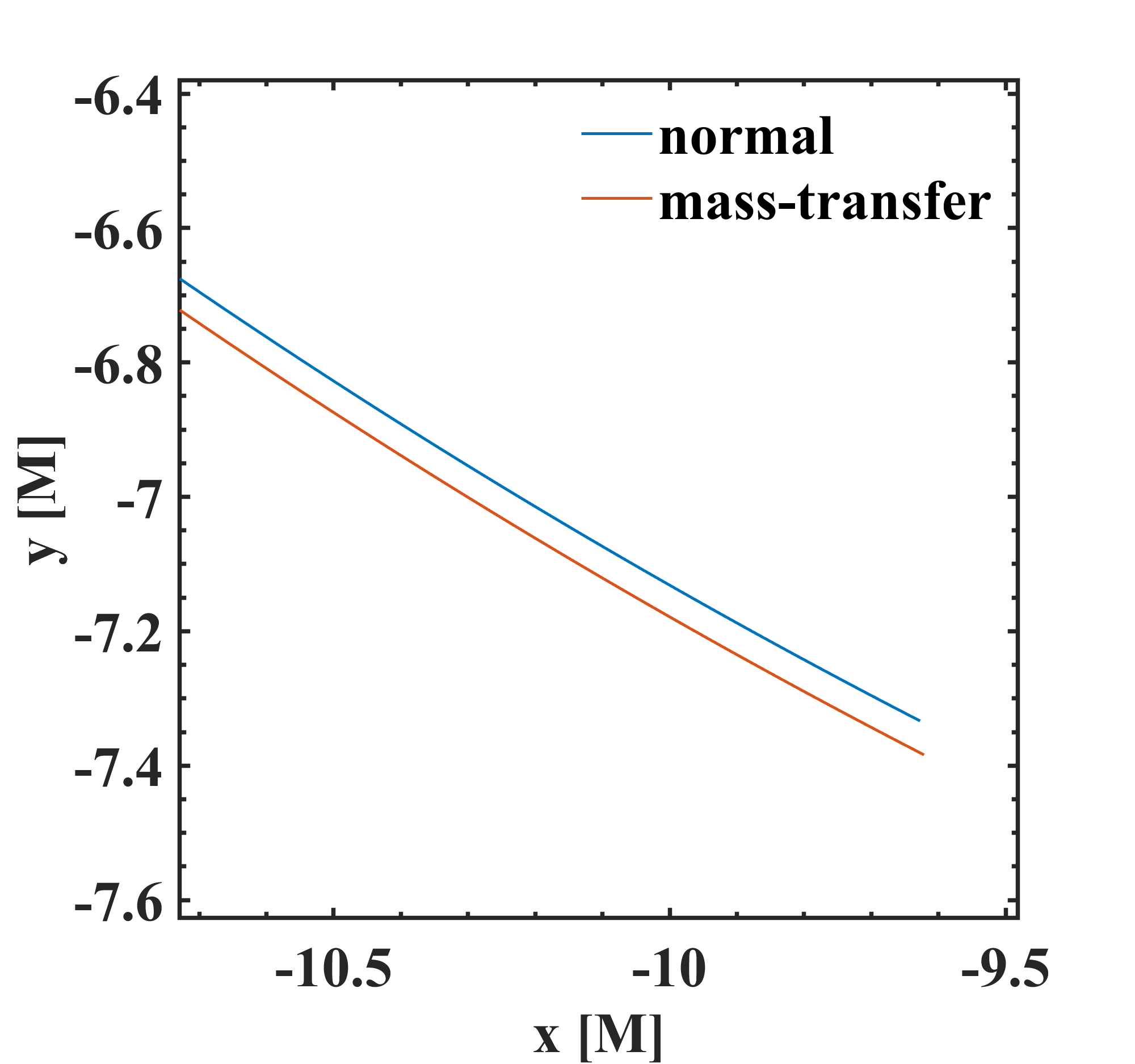}}
    \caption{\small The upper picture shows the mass transfer correction on the terminal GWF of a $(2+1.4)M_{\odot}$ NS-NS system starting with velocity $v_0=0.237c$ perpendicular to separation $r=50[M]$ and eccentricity $e=0.1$. Owing to the compactness of NSs, the difference is tiny and only visible on the tips near the merge; The bottom left figure is the corresponding orbit where we set the primary star on the origin. The difference in orbital trajectory is even smaller where the termination is zoomed in and plotted on its right, which shows that mass transfer will lead to a smaller orbital change rate in the later stage of the inspiral phase relative to no mass transfer.}
	\label{NS-NS}
\end{figure}
As shown in Fig.\ref{NS-NS}, the density of double neutron stars is close, which caused the small Roche limit, the mass transfer correction is very tiny to NS-NS systems in the inspiral phase. In the merge and ringdown phases, numerical relativity and neutron star state equations are needed to calculate, thus our main focus of this paper is on NS-WD systems.

\subsection{NS-WD System}
When the density of the WD is high enough and the temperature is cool compared to the Fermi temperature of electrons, the WD can simply be treated as a ball stably balanced by the degeneracy pressure of ironized electrons \cite{LiuLiao}. In non relativistic case, the equation of states is
\begin{equation}
    P=\frac{1}{5}(3\pi^2)^{2/3}\frac{\hbar^2}{m_e}\left(\frac{\rho}{\mu_em_N}\right)^{5/3}.
\end{equation}
This is a polytrope model of $\gamma=5/3$ and $n=3/2$ and $m_N$ is the static mass of free nucleon. We can further obtain the total mass $m_{WD}$.
\begin{equation}
    m_{WD}=\frac{1}{2}\left(\frac{3\pi}{8}\right)^{1/2}(2.714)\left(\frac{\hbar^{3/2}c^{3/2}}{m_N^2\mu_e^2G^{3/2}}\right)\left(\frac{\rho(0)}{\rho_c}\right)^{1/2},
\end{equation}
where $\rho(0)$ is a constant, $\rho_c=1.687\times10^{-18} [M]^4$ is the central density and $\mu_e$ is the mean molecular weight per electron. 

The density distribution of a WD has a sharp decay, the inner of which can be approximated to a homogeneous sphere core of $\rho_c$, and outside is its atmosphere. 

Due to the sparseness of the atmosphere, its overflow has a minimal contribution to the orbital quadruple moment \cite{Alam_2023}, and the PN method is still valid ignoring its influence. Through the above mass relationship, the expression of the radius $R_{WD}$ of the white dwarf can be obtained from the Chandrasekhar limit
\begin{equation}
    R_{WD}\simeq\frac{10^{-3}}{6.963}\left(\frac{M_{WD}}{0.7M_\odot}\right)^{-\frac{1}{3}}\left[1-\left(\frac{M_{WD}}{M_{CH}}\right)^{\frac{4}{3}}\right]^{\frac{1}{2}}\left(\frac{\mu_e}{2}\right)^{-\frac{5}{3}},
\end{equation}
where $M_{CH}$ is the Chandrasekhar mass. When the WD enters the primary Roche limit
\begin{equation}
a_{Rlof}=R_{WD}\frac{0.6q^{2/3}+\ln(1+q^{1/3})}{0.49q^{2/3}},
\end{equation}
mass overflow will occur and become the common envelope of system.
We can get the density distribution and pressure distribution of the white dwarf in the critical state of disintegration \cite{Fernández_2013,10.1093/mnras/staa507}.
\begin{equation}
P=\rho_0\frac{2Gm_{WD}}{5R_0}\left[\frac{R_0}{r}-\frac{1}{2}\left(\frac{R_0}{r\sin\theta}\right)^2-\frac{1}{2d_0}\right],
\end{equation}

\begin{equation}
\rho_0=\rho_{max}\left[\left(\frac{2H}{R_0}\right)\frac{2d_0}{d_0-1}\left(\frac{R_0}{r^2}-\frac{1}{2}\frac{R_0}{r\sin\theta}-\frac{1}{2d_0}\right)\right]^{7/2},
\end{equation}
where $R_0=a_{Rlof}/(1-q)^2$ is the radius of the center (density maximum) of the torus, $H$ is the torus scale height, $d_0=1.2,1.5,3$ is a distortion parameter which measures the internal energy content of the torus, and $\rho_{max}$ is the maximum density, in most cases we adopt $\rho_{max}=1$.

The density and pressure expressions are converted to the white dwarf centroid coordinate system, and we can obtain the 
\begin{gather}
    \sin\theta=\frac{a_{Rlof}-r_{c}\sin(\theta_{c}+\phi)}{\sqrt{a_{Rlof}^2+r_{c}^2-2r_{c}a_{Rlof}\sin(\theta_{c}+\phi)}},\\
    r=[r_c^2+a_{Rlof}^2-2r_{c}a_{Rlof}\sin(\theta_c+\phi)]^{1/2},
\end{gather}

where $r_c$ is the distance to center-of-mass of WD, $\theta_c$ is the angle between $\pmb{r}$ and $\pmb{r}_c$. So that, we can obtain the analytical solution of the WD density index in the whole inspiral process.

\subsection{Mass Transfer of Quasi-Circular Orbit}
\begin{figure}[ht!]
    \centering
    \subfigure{\includegraphics[width=1\linewidth]{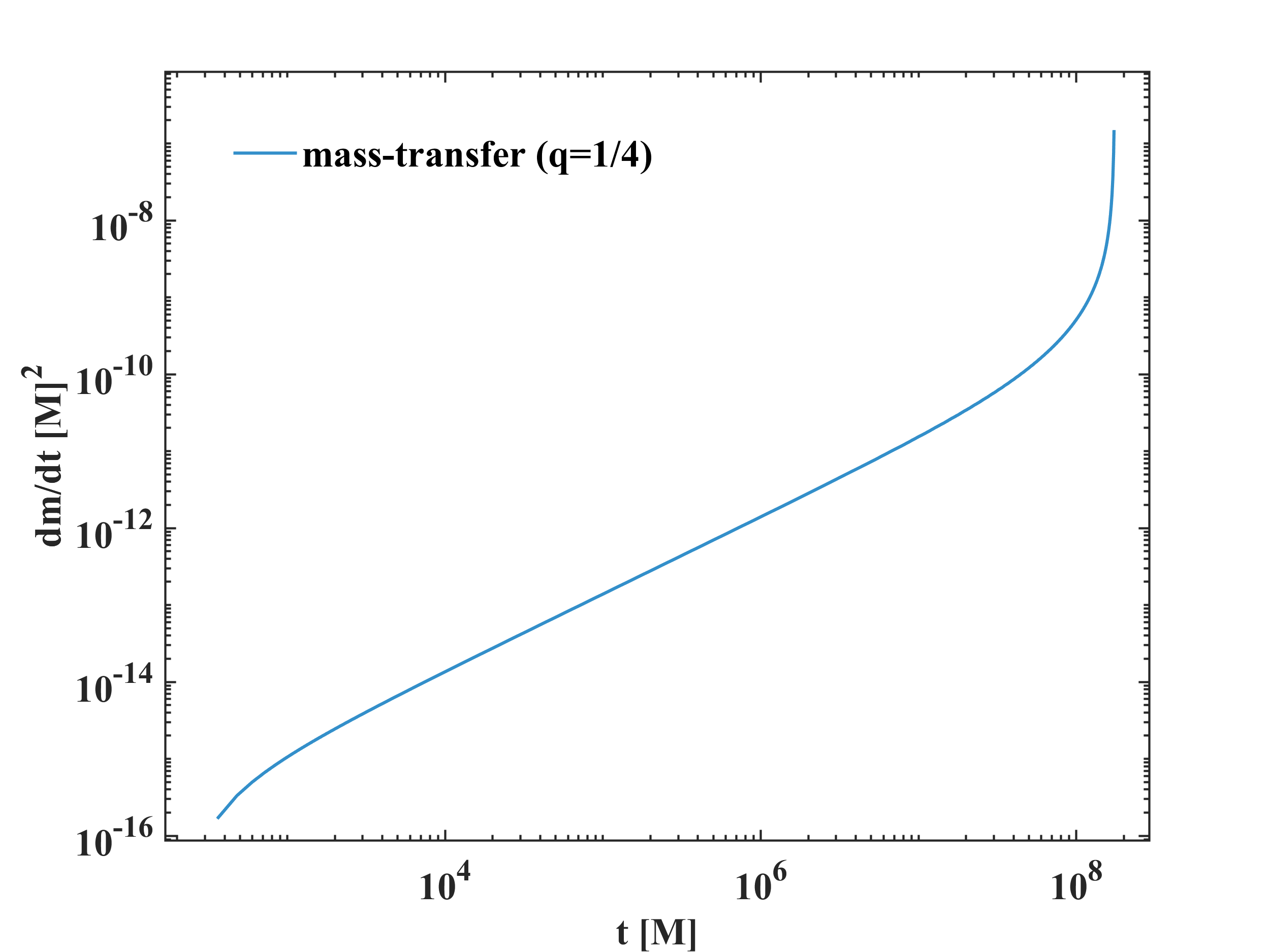}}\\
    \vspace{-0.1in}
    \subfigure{\includegraphics[width=1\linewidth]{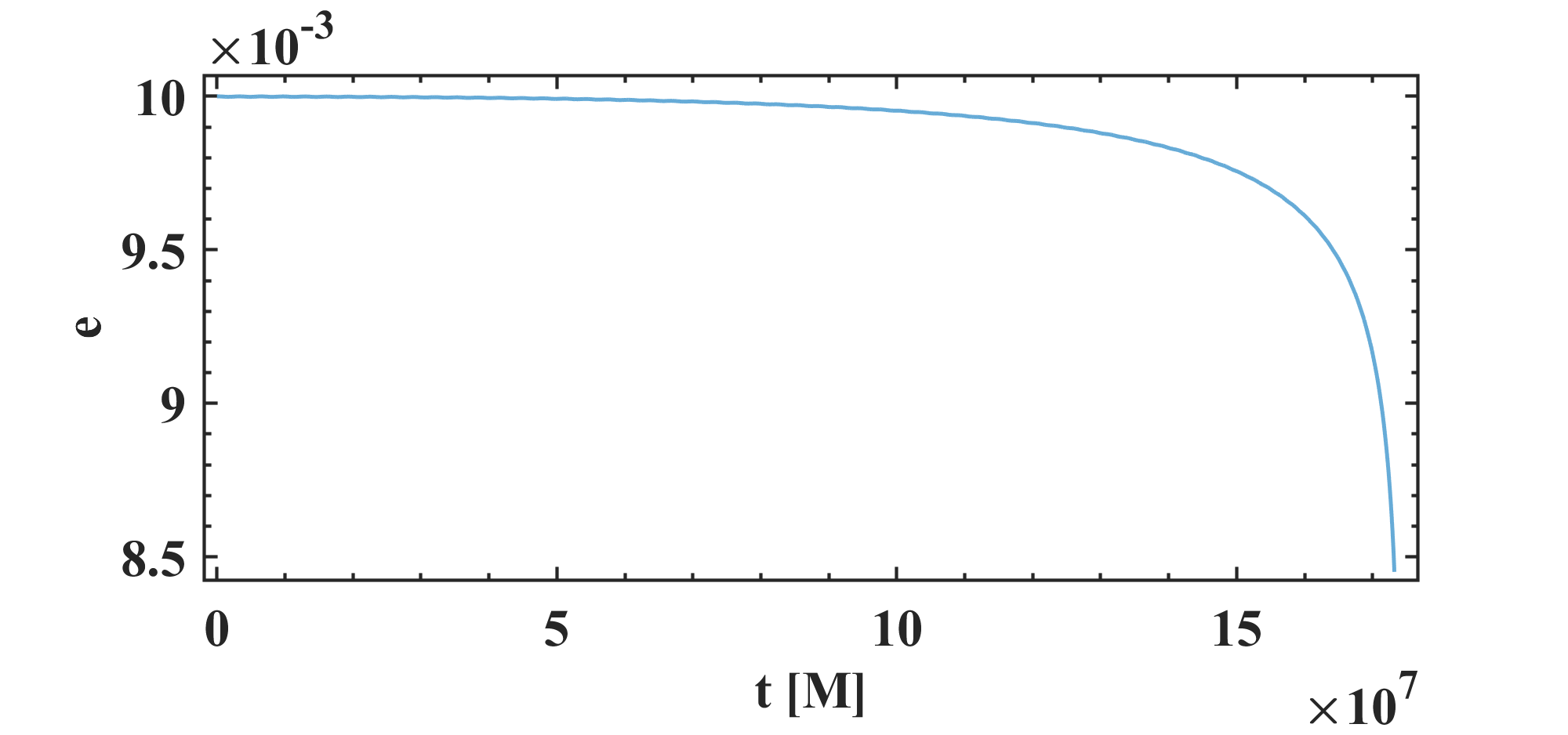}}
    \caption{\small 
    In this figure, the initial distance of the binary system is the Roche limit of NS with $e=0.01,q=1/4$. Below is an image of the  eccentricity change, which is basically stable at the initial stage of the inspiral phase, and then decreases rapidly and tends to near 0.
	\label{or_m}}
\end{figure}

In reality, when the mass of the binary is relatively large, the orbit is closer to circular of smaller eccentricity. In this orbit, the binary system is more stable and the mass transfer is weaker. In order to study the overall trend of mass transfer over time, we discuss the mass transfer with initial eccentricity $e=0.01$ in FIG.\ref{or_m}.

\begin{figure*}[ht!]
\vspace{-0.1in}
    \centering
    \subfigure{\includegraphics[width=1\linewidth]{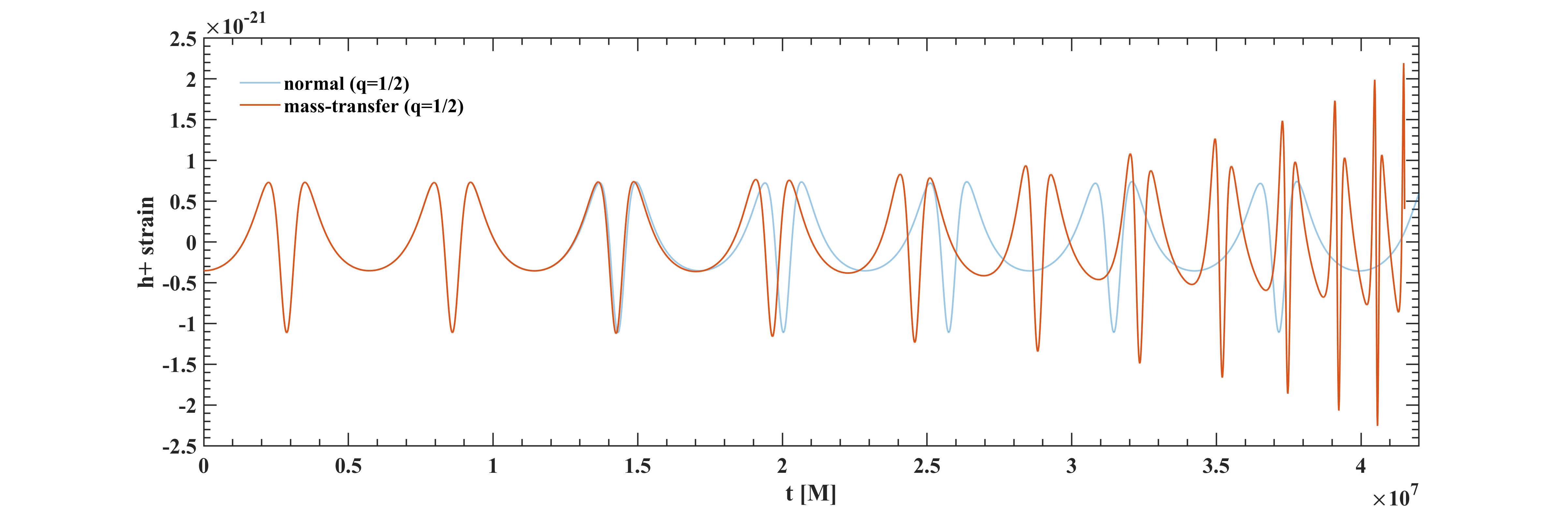}}\\
    \vspace{-0.3in}
    \subfigure{\includegraphics[width=1\linewidth]{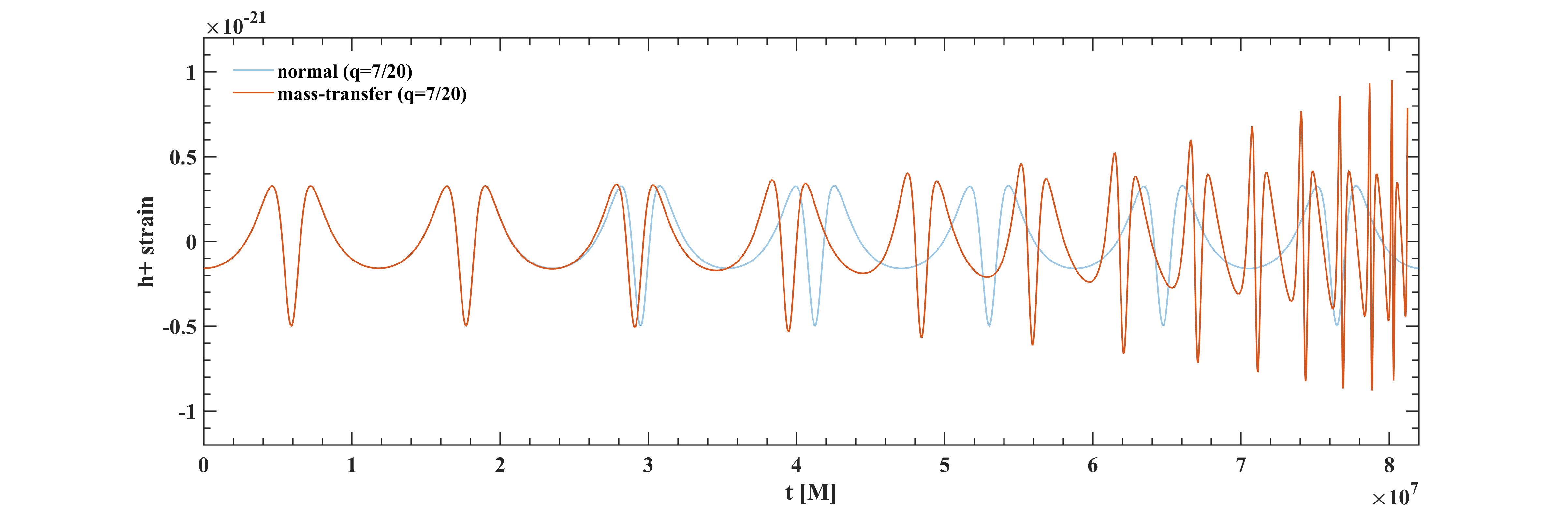}}\\
    \vspace{-0.3in}
    \subfigure{\includegraphics[width=1\linewidth]{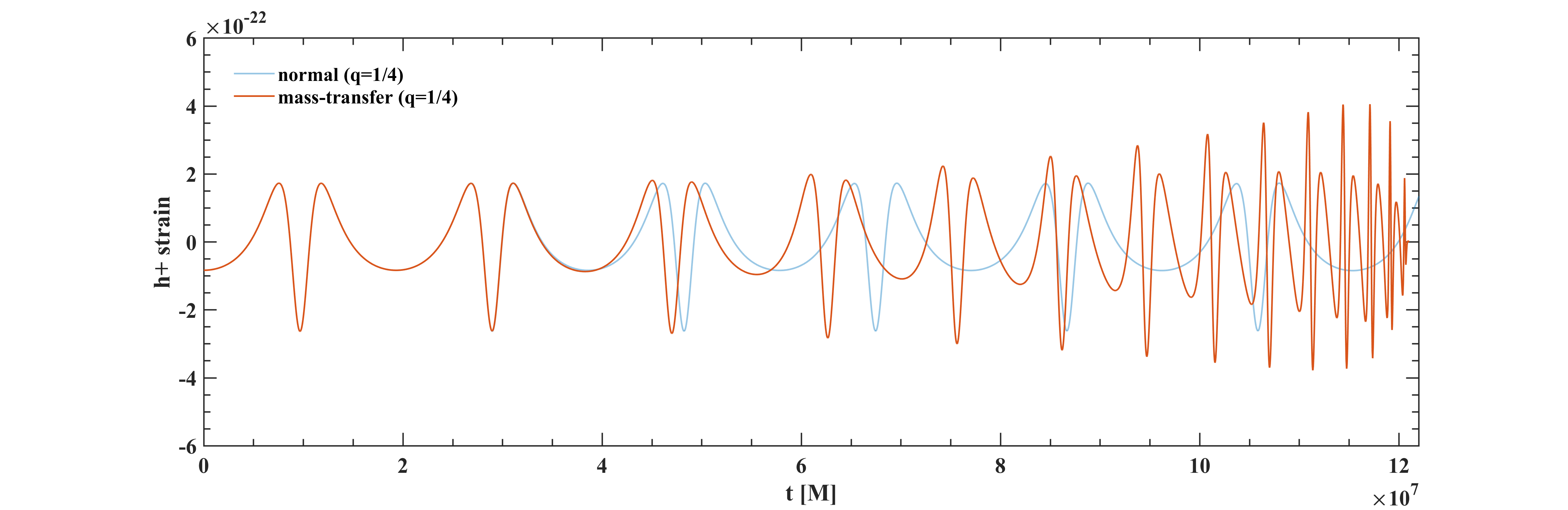}}
    \caption{\small In comparison with no mass transfer correction, the binary rotation in late stage is largely hastened by mass transfer. In addition to that, a small mass ratio shows a stronger frequency-promoting effect on the GW,  as well as stronger strain suppressing against the chirp, where the strain of the $q=1/4$ one even decays at the terminal. Detailed records are listed in TAB.\ref{WD_NS_tbl}.}
	\label{WD_NS}
 \vspace{-0.1in}
\end{figure*}

\begin{table*}[ht!]
\centering
\caption{Key parameters of our simulation, including the initial radius of WD $R_{WD}^{(i)}$, Roche limit $a_{Rlof}^{(i)}$ and relative velocity $v_0$ which is set to be Keplerian, the final mass of the NS $m^{(f)}_{NS}$, WD $m^{(f)}_{WD}$, and AD $m^{(f)}_{AD}$ and branch parameter $\kappa$ of three different initial mass ratios $q=1/2, 7/20, 1/4$.}
\renewcommand\arraystretch{1.6}
\begin{tabular*}{\hsize}{@{}@{\extracolsep{\fill}}c|cccccccc@{}}
\toprule
 \qquad NS-WD mass \qquad\qquad & $R_{WD}^{(i)} ([M]^{-1})$ & $a_{Rlof}^{(i)} ([M]^{-1})$ & $v_0 (c)$ & $m^{(f)}_{NS} ([M])$&$m^{(f)}_{WD} ([M])$ &$m^{(f)}_{AD} ([M])$&$\kappa$\;\\
\midrule
\;$(2.0+1.0) [M]\;$ & 3673.036 & 11450.037 & 0.0097 & 2.475 & 0.462 &0.063&0.454\;\\
\;$(2.0+0.7) [M]\;$ & 5239.930 & 17901.543 & 0.0073 & 2.476 & 0.174 & 0.050 &0.396\;\\
\;$(2.0+0.5) [M]\;$ & 6484.328 & 24231.317 & 0.0061 & 2.452 & 0 &0.048 &0.207\;\\
\bottomrule
\end{tabular*}
\label{WD_NS_tbl}
\end{table*}
\clearpage

Since the quasi-circular orbit rotates in the same period, the distance $r$, velocity $v$, and other orbit parameters change little, so the whole mass transfer rate function tends to be monotonous and smooth, which helps us to analyze the overall mass transfer process. In the whole process of mass transfer, the mass transfer rate rises rapidly in the first short period of time and then keeps rising gently for a long time until the later stage of evolution.

For more general cases, there will be a lot of binary systems with large eccentricity. Due to the rotation along the elliptical orbit, the WD will orbit in and out of the NS's Roche limit in each cycle. At this time, the mass transfer rate will be based on the overall trend of the original quasi-circular orbit, and the details of the oscillation changes in each period will be added. Therefore, in the following text, we mainly discuss the binary star system with eccentricity $e = 0.4$.

\subsection{Binary Systems of Different Mass Ratios}

According to our theoretical analysis, the NS-WD binary has more complicated mass transfer processes, and in FIG.\ref{WD_NS} we plot three $h_+$ waveforms of different mass ratios with initial eccentricity $e=0.4$ and distance $r_0=(1+e)/(1-e)a_{Rlof}^{(i)}$. On this initial condition, the calculation begins at the instance when the WD first time enters the NS Roche limit. Depending on the evolution, the final states can be divided into radius truncation and mass truncation. In the former condition, two stars collide when their distance is smaller than the sum of their radii, and in the latter condition, the WD loses all mass in the form of mass overflow during the rotation with no violent collision.

It can be seen in TAB.\ref{WD_NS_tbl} that the radius of the WD is inversely proportional to the mass, and its density decreases rapidly as the mass decreases. The mass transfer rate $|\delta m_c|/m_c$ of WD increases as the mass ratio decreases, and the branch parameter $\kappa$ decreases as the mass ratio decreases. When $q = 1/4$, $|\delta m_c|/m_c=1$, the WD is completely disintegrated during rotation and all its mass enters the NS or accretion disk in the form of interstellar flow. 

\begin{figure}[ht!]
    \centering
    \subfigure{\includegraphics[width=1\linewidth]{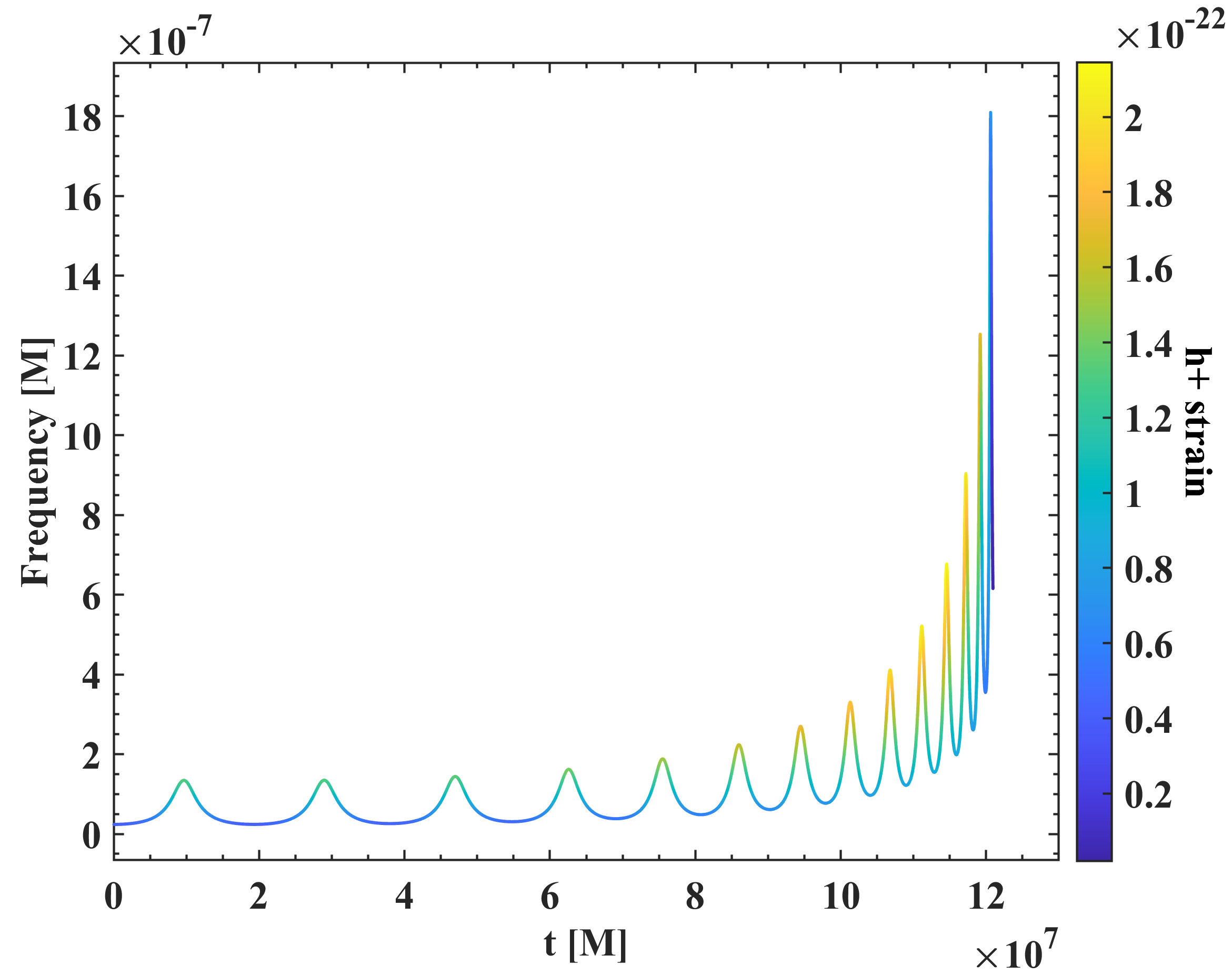}}
    \caption{\small 
    The expressive strain and frequency evolution of the $q=1/4$ one in FIG.\ref{WD_NS}. Owing to the terminal strain decay, the frequency of the main detective signal is much lower, so the extreme mass ratio ones are very similar to noises.
	\label{fts}}
\end{figure}

In order to future investigate the underlying causes of the variation in GWF, we investigate a general range of mass ratios, which are still commonly seen in actual cases \cite{PhysRevLett.126.021103,PhysRevLett.79.1186}. We plot the evolution of dynamical parameters of $(2.0+0.5)M_\odot$ to $(2.0+1.0)M_\odot$ NS-WD systems, including the radiation power, and orbital radius. The radiation power increases significantly as it approaches the merger, and we get the illustration of GW's frequency-time-strain in FIG.\ref{fts} in the cases of $q=1/4$. The main difference appears in the late inspiral phase, when $m_c$ star moves within the Roche limit of $m_p$ star, and before the merger phase.

\begin{figure}[ht!]
    \centering
    \includegraphics[width=1\linewidth]{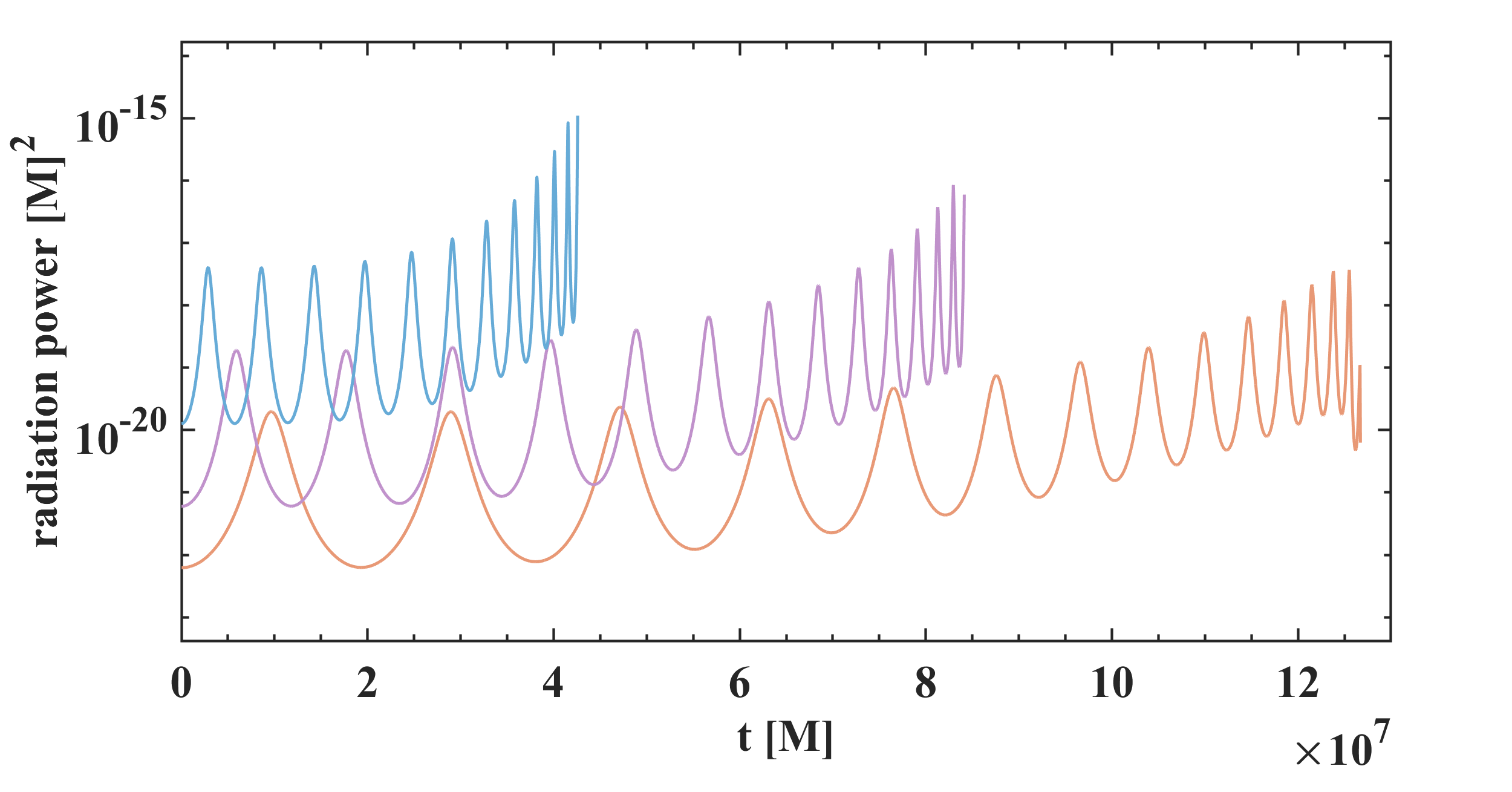}
    \caption{\small The figure shows their gravitational radiation power, which decreases rapidly as the mass ratio gets smaller. }
	\label{ep}
\end{figure}

The power of gravitational radiation is selected as an important parameter indicating the rotation of binaries, shown in FIG.\ref{ep}. As the mass ratio gets smaller, the initial distance increases and the GW radiation gets weaker, so the binary takes longer to merge. Due to the significant difference in the magnitude of gravitational radiation power of different mass ratios, we draw it on the logarithmic coordinate axis. The frequency and intensity of all three GWFs show clear chirp behavior. Particularly, in the case of the mass truncation, the gravitational radiation power decreases straightly when the WD disintegrates after the chirp.

\begin{figure}[ht!]
    \centering
    \subfigure{\includegraphics[width=1\linewidth]{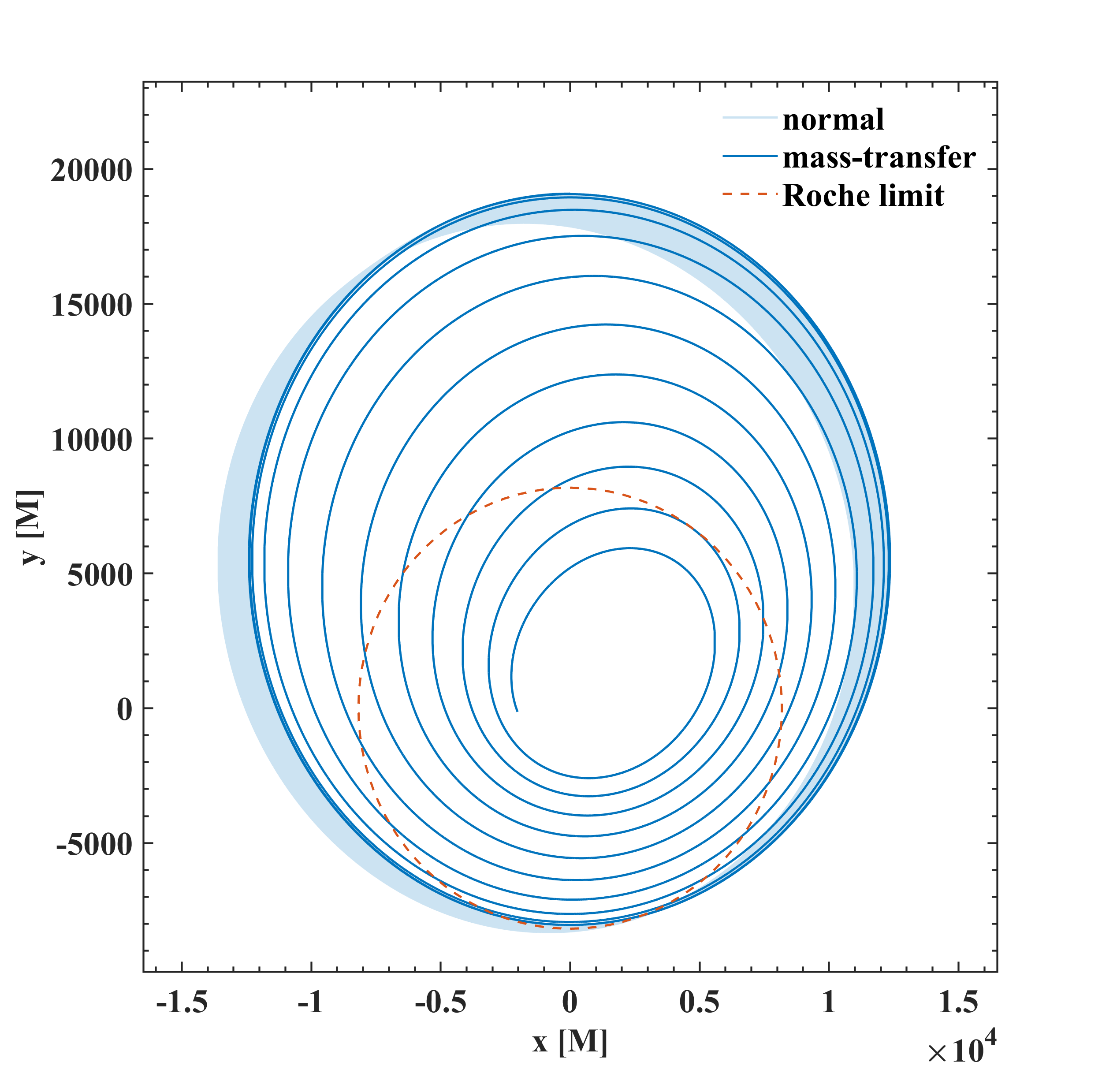}}\\
    \vspace{-0.2in}
    \subfigure{\includegraphics[width=1\linewidth]{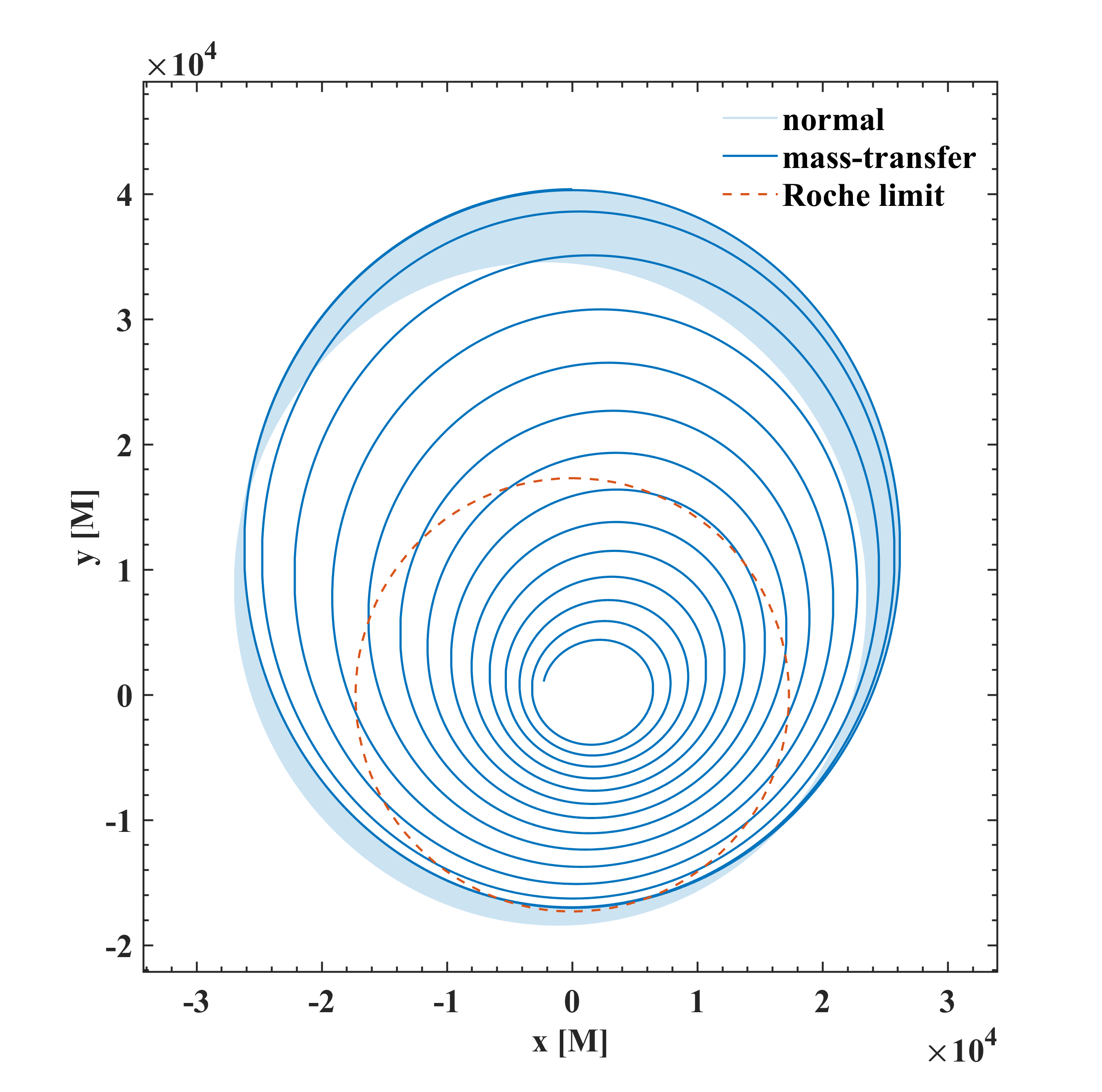}}   
    \caption{\small The upper figure illustrates the orbit of the WD of $q=1/2$ one truncated by radius, while the lower figure illustrate that of $q=1/4$ one truncated by mass. Both set the origin to be the center of NS. The orbits with no mass transfer hardly drop and are plotted as shaded bands. The circular red dotted line is the initial Roche limit of the NS.
	\label{WD_or}}
\end{figure}

In FIG.\ref{WD_or}, we draw the explicit orbits of NS-WD systems of two different mass ratios. Without the angular momentum taken away by the,  the ones without mass transfer are still rotating peripherally when the corrected ones have already merged. In $q=1/2$, the radial velocity is getting larger and larger with a precession angle. When the final radius is truncated, the binary distance $r = 2.0285\times10^3 [M]^{-1}$, which is equal to the radius of the WD. In $q=1/4$, the eccentricity gradually decreases to the quasi-circular orbit. When the final mass is truncated, the binary distance $r=5.2265\times10^3 [M]^{-1}$, and the remaining radius of the WD core  is $R_{WD}=28.7164 [M]^{-1}$. In a more extreme case, the orbit tends to be stable and takes longer to merge. 

\begin{figure}[ht!]
    \centering
    \subfigure{\includegraphics[width=1\linewidth]{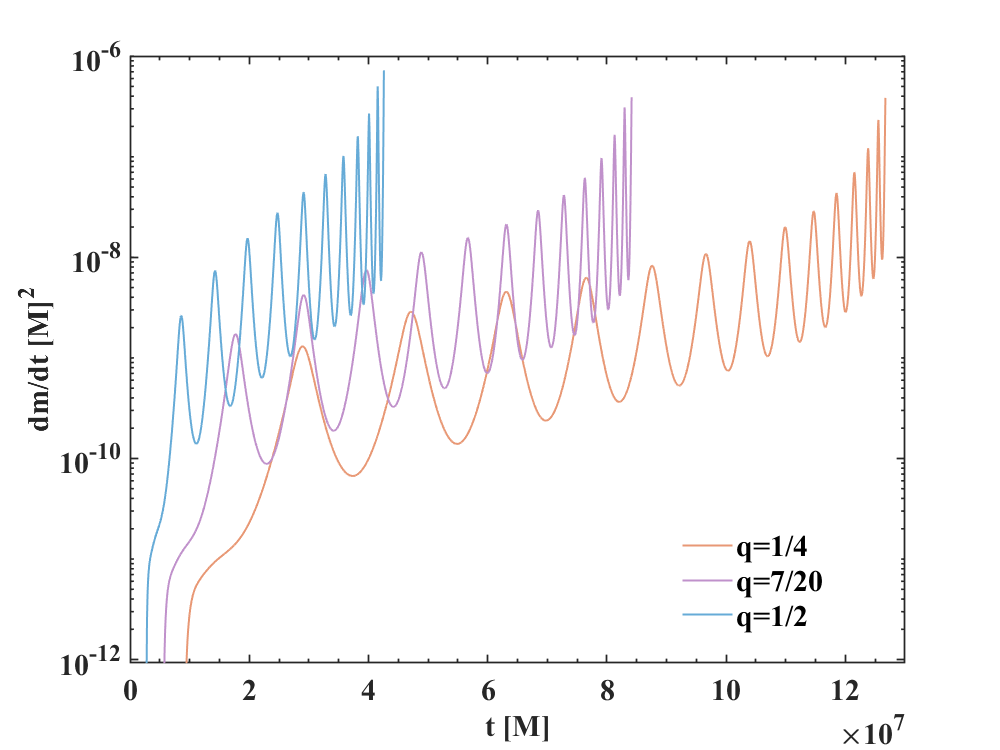}}\\
    \vspace{-0.15in}
    \subfigure{\includegraphics[width=1\linewidth]{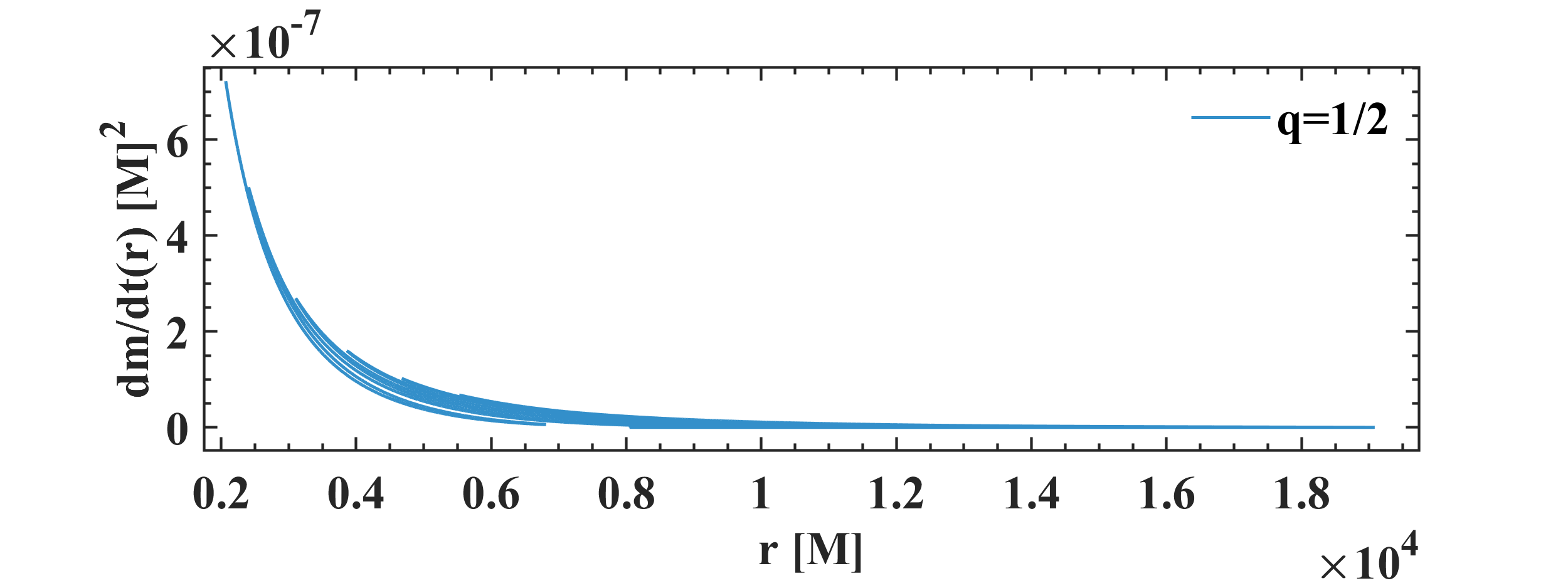}}\\  
    \vspace{-0.15in}
    \subfigure{\includegraphics[width=1\linewidth]{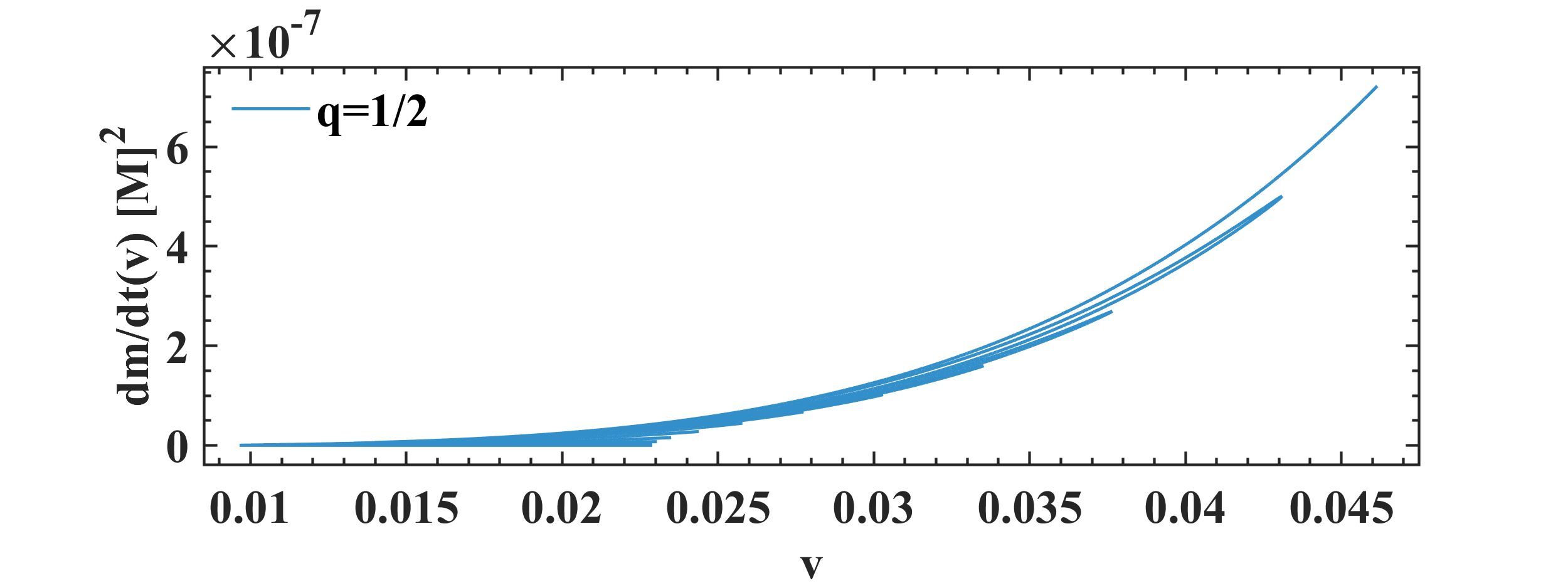}}
    \caption{\small 
    The upper figure shows the mass transfer rate of the three cases, which all increases rapidly as the distance shrinks. The oscillation is due to the competition of tidal force and centrifugal force. Take the $q=1/2$ for example, two figures below give the mass transfer ratio concerning binary distance and relative velocity in different time intervals. 
	\label{WD_dmtr}}
\end{figure}

Since $\dot{m}=\dot{m}(r,\omega)$ is a function of the angular velocity $\omega$ and the distance $r$, we plot the evolution of the mass loss rate of the WD with time and the with radius in FIG.\ref{WD_dmtr}. Considering only time, the mass transfer rate has to do with the binary distance and velocity, leading to the nonlinear result. The significant increase process of the mass transfer rate in the previous period is the stage when the outer gas overflows into the Roche limit, but the core has not yet disintegrated. When the Lagrangian point $L1$ enters within the radius of the WD, the WD core begins to tidal disrupt and oscillates periodically with the change of orbital parameters. Considering only distance $r$, the mass transfer rate increases when the separation decreases. And in \cite{10.1111/j.1365-2966.2011.19747.x} Metzger found the changing mass ratio is around $10^{-10}\thicksim10^{-7}\;[M]^2\backsimeq 10^{-4}\thicksim10^{-1}\;M_\odot/s$ for tidal disruption of a WD by a NS, consistent with our results well.

\begin{figure}[ht!]
    \centering
    \subfigure{\includegraphics[width=1\linewidth]{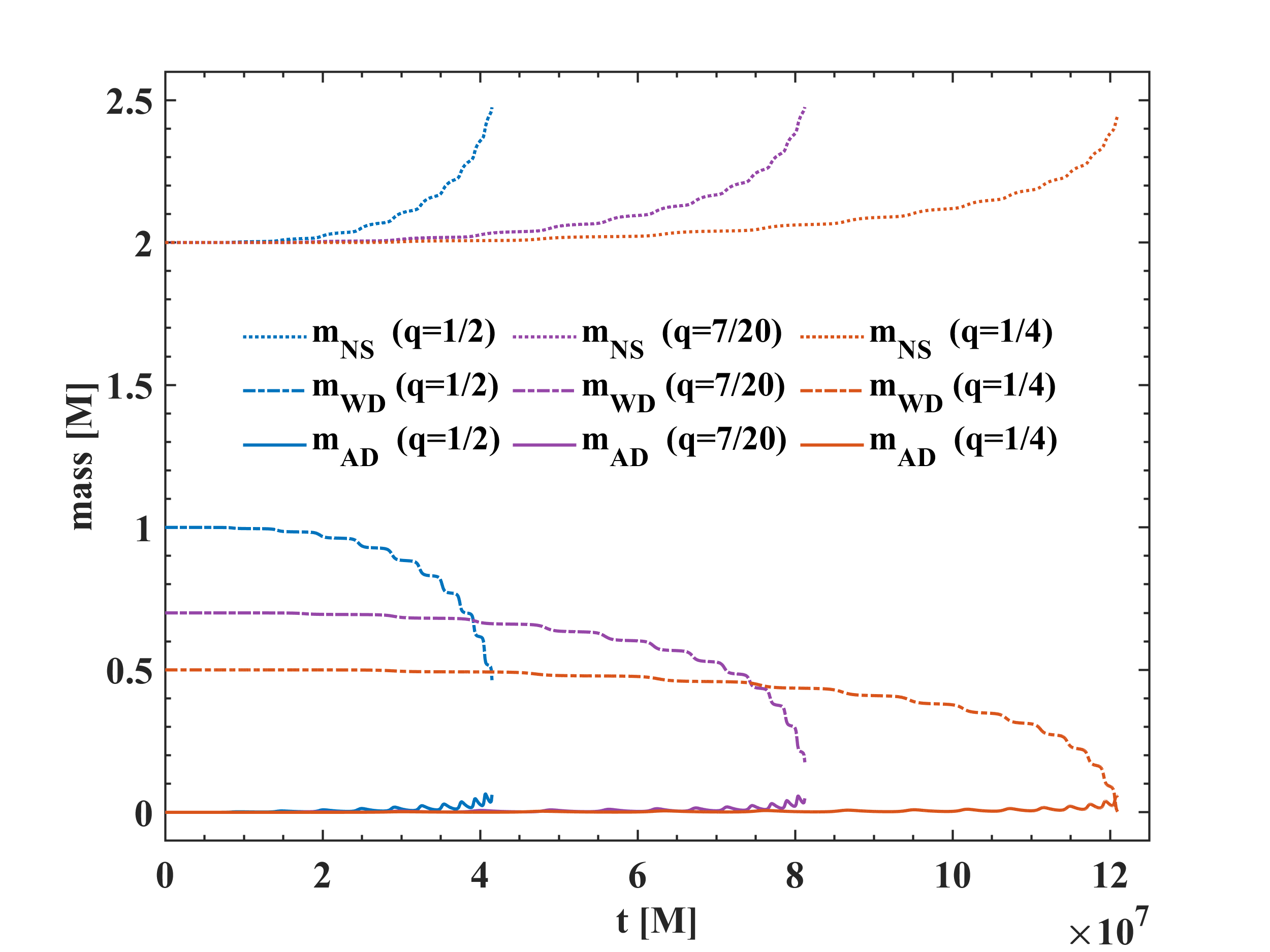}} 
    \caption{\small 
    This picture shows the NS, WD and AD mass change of the three cases respectively. Here we can see the mass of the AD is a very tiny fraction of the system, so despite that of real binary is not as symmetric as we assumed here, their contribution to the GW is still negligible.
	\label{WD_m}}
\end{figure}

\begin{figure}[ht!]
    \centering
    \subfigure{\includegraphics[width=1\linewidth]{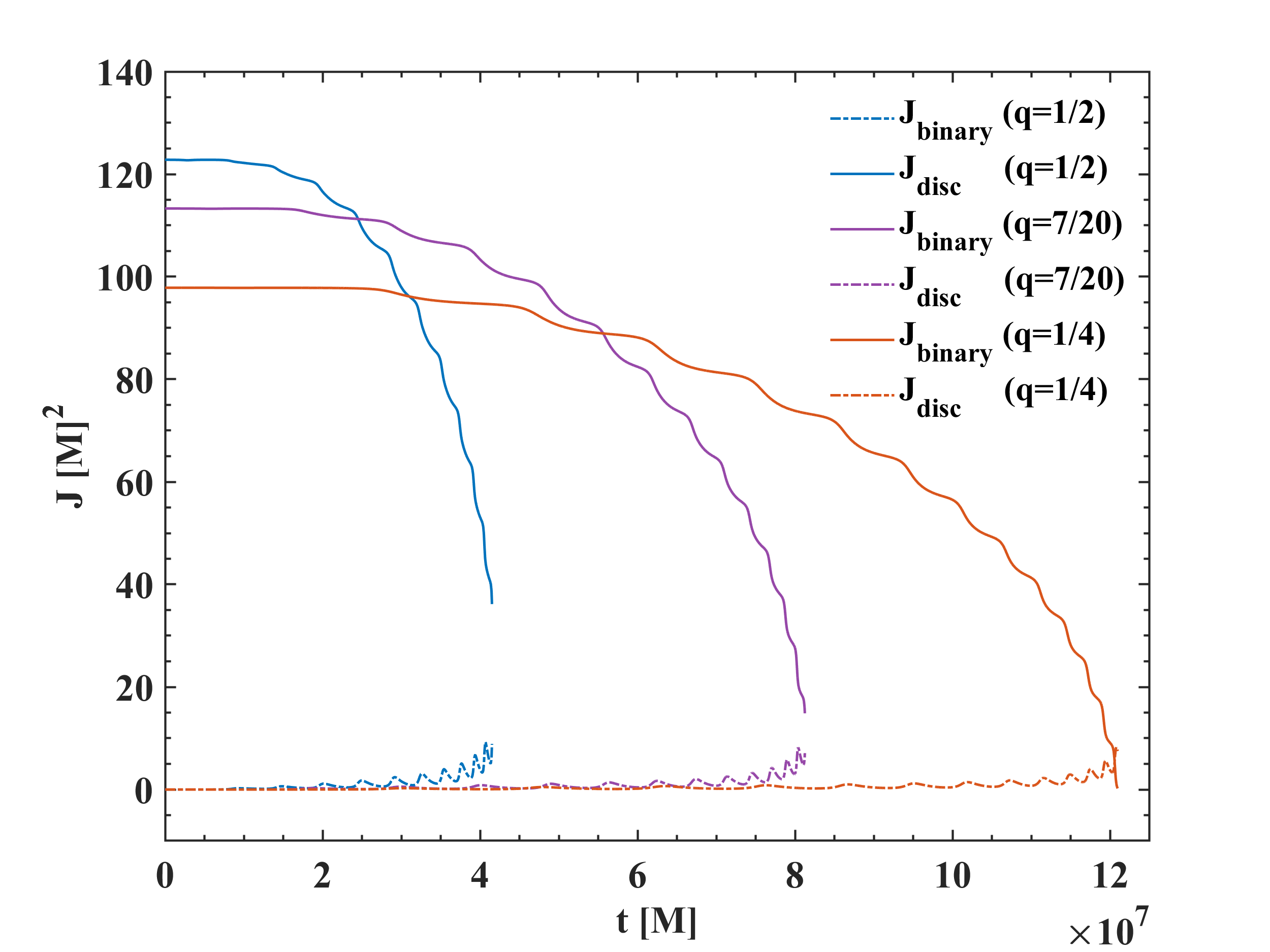}}  
    \caption{\small 
    This picture shows the corresponding angular momentum change of three cases. The stage-like behavior is inherited from the severe oscillation of the mass transfer rate, the same as the behavior of the previous picture. When $q =1/4$, due to the mass loss, the final angular momentum tends to zero.
	\label{WD_J}}
\end{figure}
\newpage
In order to get the evolution information of $m_{NS}(t)$, $m_{WD}(t)$ and $m_{AD}(t)$, we plot their curve in FIG.\ref{WD_m}. It is very characteristic that the final masses of NSs are all the same, which is caused by the variation of the lost mass of the WD $\delta m_c$ and the branch parameter $\kappa$. On the contrary, there is a significant difference in the lost mass of WDs.

We draw the angular momentum of three systems with time in FIG.\ref{WD_J}. Due to the dissipation of gravitational radiation, the systems' angular momentum is decreasing. Along with that, the orbital angular momentum is also decreasing due to the growing NS spin.

To summarize, as the mass of WD increases the binaries tend to merge more quickly leading to an increase in the changing rate of the mass quadrupole tensor, and the gravitational radiation power and GW strain during the inspiral also increase. From the discussion above we can intuitively see mass-transfer's prominent role in the orbit evolution and GWF of NS-WD inspiral, especially at the terminus close to merge.

\subsection{Matching the Template with GW170817}

\begin{figure*}[ht!]
    \centering
    \includegraphics[width=1\linewidth]{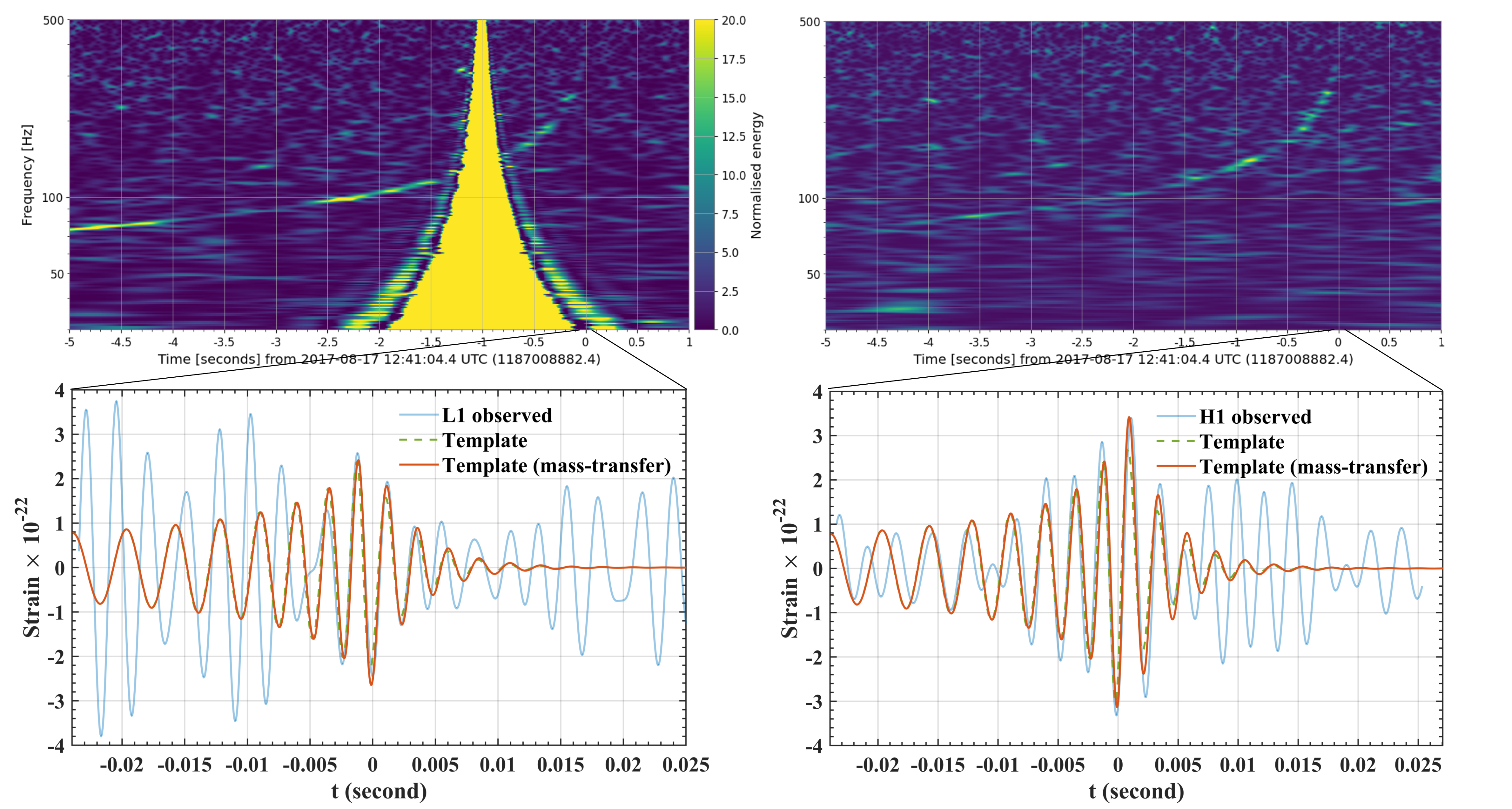}
    \caption{\small The comparisons to the GW170817 data by LIGO \cite{PhysRevLett.119.161101,PhysRevX.9.011001}, which is labeled by blue lines. The top panels are the initial data drawn in the frequency domain, where we can see the chirp rising. The left one from L1 has encountered some technical issues showing as the peak-shaped image, which leads to the failure of further processing. This is why the corresponding signal below is severely distorted, where only the merger is readable. The precise fitting parameters are listed in TAB.\ref{tbl1}}
	\label{AAA}
\end{figure*}

\begin{table*}[ht!]
\centering
\caption{The detection data and the initial parameters of numerical calculation of GW170817. $\mathcal{M}$ is the chirp mass, $M$ is the total mass, $R$ is the distance from us, $\theta$ is the viewing angle, $\chi$ is the effective spin parameter, $q$ is the mass ratio, and $e$ is the eccentricity of orbit.}
\renewcommand\arraystretch{1.6}
\begin{tabular*}{\hsize}{@{}@{\extracolsep{\fill}}c|ccccccccc@{}}
\toprule
GW170817 &$m_1(M_\odot)$&$m_2(M_\odot)$ &$\mathcal{M}(M_\odot)$ &$M(M_\odot)$&$R$(Mpc)&$\theta(^\circ)$&$\chi_{eff}$ & $q$ & $e$\;\\
\midrule
Detection data & $1.46^{+0.12}_{-0.10}$ & $1.27^{+0.09}_{-0.09}$ & $1.188^{+0.004}_{-0.002}$ & $2.74^{+0.04}_{-0.01}$ &$40^{+7}_{-15}$ & $\leqslant56$ & $0^{+0.02}_{-0.01}$ & $1.15^{+0.19}_{-0.15}$ & -\;\\
\;\;\;Setting parameters\;\;\;  & 1.46 & 1.27 & 1.185 & 2.73 & 42 & 56 & 0 & 1.15 &0.2\;\\
\bottomrule
\end{tabular*}
\label{tbl1}
\end{table*}
To verify our models, it is helpful to compare them with the actual data. Since no NS-WD signals have ever been detected, here we choose GW170817, the first NS-NS merger signal, although the mass transfer effect of the NS-NS system is not particularly obvious. This event occurred at GPS time 1187008882.43 = August 17 2017, 12:41:04.43 UTC, was widely observed by many electromagnetic bands simultaneously, and has a very high confidence level to be a NS-NS binary. Since the masses of the experimental signal are much more precise than the distance, we follow the given mass and set the distance and eccentricity to be fitting parameters. Our priority is to match the merger signal with the highest confidence, which turns out very well. The mismatch of the inspiral may come from the choice of initial velocities. Also, due to the frequency filter which would distort the low frequency inspiral, and the Livingston signal is scratched, the experimental data may not be very faithful, so this mismatch is preserved. 

As for the post-merger waveform, we can see the unusual amplitude increasing in the reference data \cite{PhysRevD.57.4566,PhysRevD.57.4535}. This results from the small mass of NS-NS binary, that this part of the signal is beyond the LIGO's sensitivity and of very low confidence level \cite{PhysRevD.101.044006}. Here we draw the black hole ringdown waveform, simply for completeness, which does not represent the real model image after the merger of NS-NS binaries. This means that when we consider the influence of the spin of the binary system in 2.5PN, the motion equation of the whole system will be more accurate.

We obtained the data from the Gravitational Wave Open Science Center website, and let the whitened data pass a low-frequency filter to reduce the background noise \cite{Abbott_2020}. We see that the binary orbiting frequency is around 400 Hz; thus, we portray the signals between [300 Hz, 500 Hz]. We set the merger time to be the origin and focus on the waveforms between [-0.023 s, 0.023 s], and the results are shown in FIG.\ref{AAA}.  

From the contrast above, our model is generally valid to reach the expectations in practical terms. Due to the high frequency of the binary neutron star system, it is beyond the detection range of the current gravitational wave detector, so the accurate waveform template of the inspiral phase is of great significance.

\section{CONCLUSIONS}
 In our research, it is found that the evolution of WDs may be divided into two cases due to different external conditions and degrees of evolution: 1. stable mass transfer; 2. white dwarfs break out due to thermonuclear reaction. This paper focuses on constructing an analytical model of stable mass transfer to correct the orbital dynamics evolution of the post-Newton method in the inspiral phase. Of course, this assumption will inevitably exhibit a certain deviation in systems characterized by violent nuclear reaction processes to which our model does not apply. In the later work, we will consider more about the equation of state and evolution of WDs and the eccentricity of orbits in detail.
 
For the tidal disruption type of mass-transfer, the overall correction to NS-NS GWs is small, while a large change occurs on the NS-WD binaries. We have theoretically derived the equation for mass-transfer by tidal gravity and then used the branch parameter $\kappa$ to calculate the various types of mass evolution in the binaries and their corresponding GWs. We set the massive NS $m_{NS}=2M_\odot$ and WD $m_{WD}=0.5\thicksim 1M_\odot$, then obtained $\kappa=0.207\thicksim 0.454$, mass-loss ratio $|\delta m_{WD}|/m_{WD}=1\thicksim 0.638$, and mass of ADs $m_{AD}=0.048\thicksim 0.063$. These results are consistent with the current observation data, and the calculation is concurrent throughout the mass ranges of WD.

Based on 2.5 PN expansion, we derived the analytical binary mass transfer equation through the Euler equation of mass and energy conservation. The equation is dominated by the atmosphere overflow or core disruption depending on the separation of stars, and the transferring rate is a function of orbital velocity and distance. We also calculated the mass received by the NS and gained a faster-merging orbit due to AD angular momentum. With the equations above, we discussed the mass transfer concerning different initial mass ratios and their corresponding GWFs, which differ obviously from those without correction and benefits for constructing templates in search of Galactic compact binaries.  

In the discussions in preceding chapters, we have assumed $v_0\ll c$. The mass velocity $v_0$ is a function of only $T_0$ and $\rho_0$. So our transferring rate is determined by the atmosphere density and its average macular mass. For more details, specific WD components and equation of state must be provided and are beyond the topic of this paper, and we shall study them in the following research.

\section{ACKNOWLEDGMENTS}

This work is partially supported by the National Key Research and Development Program of China (Grant No.2021YFC2203003). We thank Prof. Zong-Kuan Guo for insightful advice on our theoretical aspects of Post-Newtonian expansions.  

Note added. — Recently, we became aware of a manuscript that independently derives results similar to FIG.\ref{or_m} for mass transferring rate change of extremely small eccentricity \cite{Chen_2023}.

\bibliography{apssamp}

\end{document}